\documentclass[preprint,12pt]{elsarticle}

\usepackage{amsmath,amssymb}
\usepackage{booktabs}
\usepackage[a4paper,left=1.8cm,right=1.8cm,top=2.0cm,bottom=2.0cm]{geometry}
\usepackage{placeins}
\usepackage{graphicx}
\usepackage{microtype}
\usepackage{array}
\usepackage{siunitx}
\usepackage{hyperref}
\graphicspath{
{./}
{qaoa_category2_full_study_results/plots_postprocessed/}
{qaoa_category2_full_study_results/noise_low_S8192/plots_postprocessed_v2/}
{qaoa_category2_full_study_results/noise_high_S128/plots_postprocessed_v2/}
{plots_postprocessed/}
{plots_postprocessed_v2/}
}

\journal{Swarm and Evolutionary Computation}

\begin{document}

\begin{frontmatter}

\title{Adaptive Differential Evolution and Multistart Search for Noisy QAOA Optimization}

\author[inst1,inst2,inst3]{Vojt\v{e}ch Nov\'{a}k\corref{cor1}}
\ead{vojtech.novak.st1@vsb.cz}

\author[inst1,inst2,inst3]{Ivan Zelinka}
\author[inst4]{Swagatam Das}
\author[inst5]{Martin Beseda}

\cortext[cor1]{Corresponding author}

\address[inst1]{Department of Computer Science, Faculty of Electrical Engineering and Computer Science, VSB--Technical University of Ostrava, Ostrava, Czech Republic}
\address[inst2]{IT4Innovations National Supercomputing Center, VSB--Technical University of Ostrava, 708 00 Ostrava, Czech Republic}
\address[inst3]{Department of Informatics and Statistics, Marine Research Institute, Klaipeda University, Lithuania}
\address[inst4]{Electronics and Communication Sciences Unit, Indian Statistical Institute, Kolkata 700108, India}
\address[inst5]{Dipartimento di Ingegneria e Scienze dell'Informazione e Matematica, Universit\`{a} dell'Aquila, Via Vetoio, I-67010 Coppito, L'Aquila, Italy}

\begin{abstract}
We benchmark classical optimization of a fixed low-depth Quantum Approximate Optimization Algorithm (QAOA) ansatz across four cost-Hamiltonian families at $N=12$, $p=3$, and $D=6$. Ten optimizers are compared over 25 independent runs under common ceilings of 10\,000 and 30\,000 function evaluations (FEs), first with exact statevector objectives and then with two additive observation-noise levels. Exact objectives favor multistart BFGS and multistart CMA-ES. Under noisy feedback, adaptive population methods become more competitive, but the ranking depends on whether performance is measured by the best exact point visited or by the point selected from noisy observations. A targeted extension over all three pre-screened instances per family confirms this regime change while showing that named adaptive-DE winners are instance dependent: jSO-lite leads low-noise oracle search, iL-SHADE high-noise oracle search, and L-SRTDE high-noise selected solutions in the equal-instance summaries. Bootstrap analysis quantifies a non-negligible high-noise search--selection gap, and a retrospective fixed-budget verification proxy shows that reserving a small measurement budget for final re-evaluation improves selected quality across all ten methods at 30\,000 FEs. A supplementary structure-aware study further shows that QAOA cross-depth restriction and continuous basin refinement are more useful in these conditions than standalone Monte Carlo tree-search selection. Overall, optimizer choice depends jointly on landscape structure, observation noise, and final-point identification.
\end{abstract}


\end{frontmatter}

\section{Introduction}

The Quantum Approximate Optimization Algorithm (QAOA) is a hybrid variational method designed to find approximate solutions to combinatorial optimization problems by encoding cost functions into diagonal Hamiltonians \cite{Farhi2014,Zhou2020}. Like other variational quantum algorithms, including VQE \cite{Cerezo2021Review,Tilly2022,beseda2024state,illesova2025transformation}, QAOA couples a parameterized quantum circuit to a classical optimizer that iteratively updates variational angles from evaluated expectation values. \cite{illesova2025classical,novak2026quantum} Realizing the practical potential of QAOA, however, requires overcoming substantial classical trainability bottlenecks. At finite circuit depth, optimization performance can be strongly constrained by the basin structure of the variational energy landscape, where non-convexity, basin multiplicity, and small near-optimal catchment regions can impede local optimization well before the expressibility limits of the ansatz are reached \cite{Willsch2020,SackSerbyn2021,BoyWales2024}. Consequently, local traps and search geometry constitute a distinct trainability bottleneck alongside gradient concentration and barren-plateau phenomena \cite{AnschuetzKiani2022,Nemkov2025,Novak2026Landscape,Novak2026GlobalSearch}.

In realistic execution, expectation values must be estimated from a finite number of projective measurements, introducing statistical uncertainty into the objective \cite{Scriva2024}. On noisy intermediate-scale quantum (NISQ) processors \cite{Preskill2018}, this uncertainty is compounded by device-level effects including decoherence, gate infidelities, and measurement readout errors \cite{Kandala2017,Wang2021NoiseBP}. Such effects can suppress gradients and alter the effective optimization landscape \cite{Wang2021NoiseBP}. To isolate classical search behavior from hardware-specific error channels, we therefore study exact statevector objectives together with controlled additive Gaussian observation noise used as a simplified proxy for finite-shot uncertainty.

Under noisy feedback, classical optimizer choice becomes substantially more difficult. Two common reference strategies are Covariance Matrix Adaptation Evolution Strategy (CMA-ES) and Simultaneous Perturbation Stochastic Approximation (SPSA). CMA-ES adapts a population covariance model without requiring explicit gradients, whereas SPSA estimates a stochastic search direction from two simultaneous perturbations per iteration \cite{Hansen2006,Spall1992}. In variational quantum algorithms, Bonet-Monroig et al.\ showed that appropriate tuning can make CMA-ES competitive with, and in some cases superior to, SPSA under sampling noise \cite{BonetMonroig2023}. QAOA studies have likewise found that SPSA can remain competitive under realistic device noise \cite{PellowJarman2024}. More broadly, variational-circuit benchmarks indicate that optimizer performance depends strongly on the assumed noise model, circuit or ansatz structure, and optimization protocol \cite{Lavrijsen2020,FernandezPendas2022,Jones2025,Scriva2024}. Related studies further emphasize the roles of optimization dimension, landscape structure, statistical reliability, and optimizer configuration in determining trainability \cite{Novak2025NoisyLandscapes,Illesova2025VHA,Novak2025Reliable,illesova2025qmetric}.

A complementary family of derivative-free global optimizers is Differential Evolution (DE), particularly the adaptive success-history variants developed through the IEEE Congress on Evolutionary Computation (CEC) single-objective competitions. Following the introduction and strong CEC performance of L-SHADE, this lineage produced methods such as iL-SHADE and jSO, with later variants combining success-history adaptation, population-size reduction, and increasingly specialized mutation and crossover mechanisms \cite{TanabeFukunaga2014,Brest2016,Brest2017,Stanovov2024,Novak2026LinearProposal}. A longitudinal analysis of the CEC competitions relates this development to increasingly non-separable and rotated benchmark functions and discusses parallels with variational quantum optimization \cite{Novak2026CEC}. A recent survey of SHADE/L-SHADE variants further shows that the preferred algorithm can depend strongly on problem dimension, landscape difficulty, and evaluation budget \cite{Piotrowski2026}. Differential Evolution has also been applied directly to variational quantum algorithms as a global strategy for escaping poor local basins \cite{Failde2023}. These observations motivate testing CEC-derived adaptive optimizers on controlled QAOA landscapes rather than assuming that performance on classical benchmark suites transfers uniformly to variational quantum optimization.

Alternative global-search approaches have also been developed specifically for noisy variational objectives. Adaptive Bayesian optimization has, for example, been investigated for QAOA under exact, finite-shot, and hardware-noisy evaluation conditions \cite{Cheng2024DARBO}, while related statistical approaches emphasize the importance of uncertainty-aware evaluation and comparison in variational optimization \cite{Illesova2025Statistical, illesova2025complementarity,illesova2026importance}. Together with evolutionary methods, these studies reinforce the broader point that optimizer performance cannot be separated from the statistical properties of the objective evaluations.

Beyond unstructured black-box optimization, QAOA parameter search can exploit structure across circuit depths. Agirre et al.\ formulate QAOA parameter selection as a Monte Carlo tree-search (MCTS) problem with iterative search-space restriction, using solutions obtained at shallower depths to constrain subsequent searches \cite{Agirre2025MCTS}. More generally, parameter regularity, cross-depth interpolation, and parameter-transfer strategies have been investigated as means of exploiting structure in variational landscapes \cite{Mele2022,Lyngfelt2025Transfer,Bezdek2025ClassicalOptimization}. MCTS operates over a discretized sequential decision tree rather than directly over the continuous parameter landscape, providing a qualitatively different search mechanism from DE, CMA-ES, and local optimization. We therefore include a supplementary structure-aware comparison that separates the effects of tree search, depth-informed domain restriction, and continuous local refinement from the primary continuous-optimizer panel.

To evaluate these optimization strategies across qualitatively different energy landscapes, we benchmark a fixed-depth ($p=3$, $D=6$) shared QAOA ansatz on four combinatorial and spin-glass Hamiltonian families. A random 3-regular Max-Cut problem provides a sparse, uniform, two-local baseline without bond disorder. Sparse geometric frustration is introduced through the two-dimensional Edwards--Anderson (EA) nearest-neighbor spin glass with bimodal couplings $J_{ij}\in\{-1,+1\}$ \cite{EdwardsAnderson1975}. Higher-order interactions are represented by a diluted three-spin glass defined on a sparse three-uniform random hypergraph with couplings $J_{ijk}\in\{-1,+1\}$ \cite{GrossMezard1984,FranzParisiTriton1999}. Finally, dense all-to-all interactions are provided by the Sherrington--Kirkpatrick (SK) model, whose thermodynamic solution is characterized by a complex free-energy landscape and replica-symmetry breaking \cite{SherringtonKirkpatrick1975,Farhi2022SKInfinite}. These families span different interaction localities, sparsities, graph geometries, and frustration structures while keeping the classical optimization dimension fixed at $D=6$. Across these landscapes, we evaluate ten optimizers spanning gradient-based local search, stochastic approximation, covariance adaptation, swarm search, and several adaptive DE variants.

We compare these methods under common function-evaluation ceilings using deterministic statevector objectives and two Gaussian observation-noise regimes. Under noisy feedback, we explicitly distinguish search quality, measured by the best exact point visited, from selection quality, measured by the exact objective value of the point chosen from the noisy query history. We additionally perform a targeted multi-instance robustness extension for five leading methods to examine whether optimizer regimes persist beyond the originally selected representative of each Hamiltonian family. The aim is not to identify a universally superior optimizer, but to determine how landscape structure, observation noise, evaluation budget, and final-point identification jointly govern optimizer performance.

\section{Methods}

\subsection{Benchmark construction}

All primary benchmark conditions use standard shared QAOA,
\begin{equation}
 |\psi(\boldsymbol{\gamma},\boldsymbol{\beta})\rangle
 =
 \prod_{\ell=1}^{p}
 \exp\!\left(-i\beta_\ell\sum_{i=1}^{N}X_i\right)
 \exp\!\left(-i\gamma_\ell H_C\right)
 |+\rangle^{\otimes N}.
\end{equation}
The primary depth is fixed to
\begin{equation}
    p=3,\qquad D=2p=6.
\end{equation}
Thus the continuous dimension is independent of system size and of the number of
Hamiltonian terms. The benchmark coordinate is
$\boldsymbol{\phi}\in[-\pi,\pi)^D$. Cost angles are used directly,
$\gamma_\ell=\phi_{\gamma_\ell}$, while mixer coordinates are mapped as
$\beta_\ell=\phi_{\beta_\ell}/2$, so that the common search box spans one full
period of the mixer angle.

All objectives are evaluated deterministically by exact statevector simulation.
The cost Hamiltonians are diagonal in the computational basis, allowing the
cost layer to be applied as a phase operation while the mixer is implemented by
single-qubit $X$ rotations.

\paragraph{Hamiltonian families.}

The main study contains four families.

\paragraph{3-regular Max-Cut baseline.}
For a random 3-regular graph $G=(V,E)$,
\begin{equation}
    H_{\mathrm{MC}}=\sum_{(i,j)\in E} Z_i Z_j .
\end{equation}
Minimizing this Hamiltonian is affine-equivalent to maximizing the usual
unweighted Max-Cut objective. It provides a canonical sparse, uniform,
two-local baseline.

\paragraph{Two-dimensional Edwards--Anderson model.}
On an open near-square lattice,
\begin{equation}
    H_{\mathrm{EA}}
    =
    -\sum_{\langle i,j\rangle} J_{ij}Z_iZ_j,
    \qquad J_{ij}\in\{-1,+1\}.
\end{equation}
This introduces sparse geometric frustration while retaining two-local
interactions and low term count.

\paragraph{Diluted three-spin glass.}
A sparse random three-uniform hypergraph is used,
\begin{equation}
    H_{3}
    =
    -\sum_{(i,j,k)\in\mathcal{E}_3}
    J_{ijk} Z_iZ_jZ_k,
    \qquad J_{ijk}\in\{-1,+1\}.
\end{equation}
The number of hyperedges is approximately $1.5N$, giving a sparse higher-order
interaction model rather than a fully connected $p$-spin Hamiltonian.

\paragraph{Sherrington--Kirkpatrick model.}
The dense mean-field model is
\begin{equation}
    H_{\mathrm{SK}}
    =
    -\sum_{i<j}J_{ij}Z_iZ_j,
    \qquad J_{ij}\in\{-1,+1\}.
\end{equation}
Unit couplings are used in the implementation. To make the relation to the conventional SK normalization explicit, let
$H_{\rm SK}^{(\rm norm)}=H_{\rm SK}/\sqrt{N}$. The corresponding QAOA cost layer is
\begin{equation}
U_C(\gamma')=\exp\!\left[-i\gamma' H_{\rm SK}^{(\rm norm)}\right]
=\exp\!\left[-i(\gamma'/\sqrt{N})H_{\rm SK}\right].
\end{equation}
Thus our unnormalized coordinate $\gamma\in[-\pi,\pi)$ is equivalent to the normalized-Hamiltonian coordinate $\gamma'=\sqrt{N}\,\gamma\in[-\pi\sqrt{N},\pi\sqrt{N})$. We keep unit integer couplings so that all benchmarked Hamiltonians share the same numerical $\gamma$ box; this is a coordinate convention rather than a claim that the normalized and unnormalized Hamiltonians have the same angle period.

A fully connected three-spin ferromagnet is retained only as an optional
low-dimensional diagnostic control and is excluded from the default 25-run
benchmark because its $\binom{N}{3}$ cost terms are inconsistent with the
low-overhead design goal.

\paragraph{Instance screen.}

Before optimizer benchmarking, each random family is screened at
\begin{equation}
    N\in\{6,8,10,12,14\},\qquad p=3,
\end{equation}
using three independently generated candidate instances per $(\text{family},N)$.
For each candidate, 24 random-start BFGS quenches are used to probe the local
basin structure. Additional diagnostics include 24 random gradient probes,
10 random one-dimensional parameter scans, and Hessian characterization at
low-energy representatives.

The principal screen metrics are:
\begin{itemize}
    \item the number $K_E$ of distinct quench endpoint energy levels;
    \item the near-best catchment fraction
    $p_{0.01}$, the fraction of quenches ending within $10^{-2}$ of the best
    sampled normalized objective;
    \item the median number of local minima on random parameter-space lines;
    \item the active Hessian condition number $\kappa_{\mathrm{act}}$;
    \item the random-point gradient scale $g_{\mathrm{RMS}}$.
\end{itemize}

To avoid selecting pathological ``hardest'' disorder realizations, one
representative is chosen near the median of a composite ruggedness score formed
from within-family ranks of $K_E$, random-line minima, $\log\kappa_{\rm act}$,
and $-p_{0.01}$. Candidates with non-negligible random gradients are preferred.
The primary optimizer benchmark then uses the selected $N=12$ representative
from each family.

A separate depth control at $N=12$ evaluates $p\in\{2,3,4\}$ using the same
selected physical instances. This control is intended to test whether the
ruggedness observed at $p=3$ is already present at lower depth and how it changes
with one additional QAOA layer; it is not multiplied into the main 25-run
optimizer matrix.

\subsection{Performance measures}

The exact computational-basis spectrum is enumerated for the studied system
sizes, yielding the physical ground-state energy $E_0$ and spectral span
\begin{equation}
    \Delta_{\mathrm{spec}}=E_{\max}-E_0.
\end{equation}
A variational reference $E_{\mathrm{ref}}$ is constructed independently using
128 random-start local quenches, combined with the best point from the landscape
screen. If the subsequent benchmark discovers a lower variational energy, the
reference is updated to the best observed value.

We distinguish
\begin{align}
    \varepsilon_{\mathrm{phys}}
    &=
    \frac{E_{\mathrm{best}}-E_0}{\Delta_{\mathrm{spec}}},
    \\
    \varepsilon_{\mathrm{opt}}
    &=
    \frac{E_{\mathrm{best}}-E_{\mathrm{ref}}}
    {\Delta_{\mathrm{spec}}},
    \\
    \varepsilon_{\mathrm{var}}
    &=
    \frac{E_{\mathrm{ref}}-E_0}{\Delta_{\mathrm{spec}}}.
\end{align}
This separates classical optimization failure from the finite-depth variational
limitation of the QAOA ansatz.

\subsection{Optimization protocol}

The main comparison uses function-evaluation budgets of
\begin{equation}
    B\in\{10\,000,30\,000\}.
\end{equation}
The $10\,000$- and $30\,000$-FE conditions are independent runs rather than checkpoints of one trajectory: the budget enters the deterministic seed construction, so each budget starts from a newly generated stochastic run. The common objective wrapper counts every objective call, including population initialization, numerical finite-difference evaluations, and line-search evaluations. The nominal budget is a ceiling rather than a forced evaluation count. Budget-driven population and multistart methods normally consume the full allowance, whereas single-start BFGS and CMA-ES may terminate earlier through their native stopping criteria. Run-level FE usage is retained explicitly and audited in Appendix~\ref{app:hyperparameters}; this distinction is important when interpreting single-start versus restart-based baselines.

The full experiment uses 25 independent runs per optimizer and condition. The optimizer panel is intentionally heterogeneous. It contains BFGS and FE-matched multistart BFGS (MS-BFGS); SPSA \cite{Spall1992}; CMA-ES \cite{Hansen2006} and FE-matched multistart CMA-ES (MS-CMA-ES); iSOMA \cite{Diep2022}; SciPy differential evolution \cite{StornPrice1997}; iL-SHADE \cite{Brest2016}; L-SRTDE \cite{Stanovov2024}; and jSO-lite, adapted from jSO \cite{Brest2017}.

The configurations are fixed globally before inspecting optimizer outcomes; no model-specific or noise-specific hyperparameter tuning is performed. Where a mature library or reference implementation exposes a standard configuration, we retain that behavior as far as the common bounded domain and FE accounting allow. Hand-set values are used only to define a reproducible baseline or to adapt a long-horizon CEC implementation to the much shorter $10\,000$--$30\,000$ FE regime. This choice deliberately measures transfer of established optimizer configurations rather than the outcome of a separate hyperparameter-optimization campaign. Appendix~\ref{app:hyperparameters} records the rationale and exact settings.

The evolutionary group was selected to span different levels of adaptation. SciPy-DE provides a comparatively static DE baseline. iL-SHADE belongs to the success-history lineage and combines adaptive control-parameter memories with population-size reduction. jSO extends this lineage with stronger exploitation mechanisms; because the present FE budgets are much smaller than standard CEC horizons, we use the simplified jSO-lite configuration described in Appendix~\ref{app:hyperparameters}. L-SRTDE represents a recent CEC-2024 success-rate-based DE design. iSOMA provides a non-DE population baseline based on self-organizing migration and a narrowing search-space strategy \cite{Diep2022,Novak2026LinearProposal}. This set is motivated by the sustained CEC performance of adaptive DE families, but the benchmark does not assume that CEC ranking transfers directly to QAOA \cite{Novak2026CEC,Piotrowski2026}. This caution is consistent with broader benchmarking work showing that algorithm conclusions can depend materially on the benchmark set and comparison protocol \cite{Piotrowski2023,LaTorre2021}.

Population size is not artificially equalized across algorithms; it is treated as part of each fixed optimizer configuration, while the controlled resource is a common FE ceiling. This distinction is material for the adaptive DE baselines. At $D=6$, iL-SHADE starts from $NP_{\rm init}=4D=24$, whereas the L-SRTDE configuration follows the CEC reference scaling $NP_{\rm init}=20D=120$ before linear population reduction. The latter therefore begins with a substantially broader population but correspondingly spends more of a fixed FE budget per early generation. Because population size can interact with noisy selection, this implementation difference is reported explicitly and considered as a possible contributor---not a demonstrated cause---of the noisy-objective results.

CMA-ES and SPSA are included as established noisy-optimization references with different information use. CMA-ES ranks a population and adapts a full covariance matrix and global step size. SPSA uses two function evaluations to form a simultaneous finite-difference direction, so its per-iteration evaluation cost does not scale with $D$. Their behavior can therefore differ sharply when the landscape is multimodal or when objective noise is large relative to the local change produced by one update \cite{BonetMonroig2023}.

Single-start BFGS and CMA-ES retain their native stopping criteria. Their
multistart counterparts repeatedly launch independent runs until the common FE
budget is exhausted, with the first initialization matched to the corresponding
single-start method. This separates algorithmic early termination from the
benefit of restart-based basin coverage.

A population cap of 150 is imposed wherever the implementation exposes a
population-size control. At the default $D=6$ this cap is inactive, but it
prevents a future high-dimensional extension from spending most of the FE
allowance on initialization.

The statistical analysis respects the mixed dependence structure of the benchmark. BFGS versus MS-BFGS and CMA-ES versus MS-CMA-ES are deliberately paired within replicate because the first initialization is matched; these planned restart comparisons use two-sided Wilcoxon signed-rank tests with Holm correction across the paired family--budget tests. The remaining optimizer summaries are reported through medians, interquartile ranges, target-hit probabilities, and condition-level ranks. We do not use a single all-method Kruskal--Wallis test as confirmatory evidence because the optimizer panel mixes paired and unpaired samples. For the noisy search--selection gap, 95\% percentile intervals are obtained by bootstrap resampling optimizer restarts within each fixed family, recomputing the family median, and then taking the median over the four benchmark families. These intervals quantify restart uncertainty conditional on the chosen families rather than population-of-families uncertainty. Detailed restart statistics are reported in Appendix~\ref{app:statistics}.

\paragraph{Noisy objectives.}

Two noisy extensions reuse the same four $N=12$, $p=3$ instances and the same
FE budgets. The value returned to the optimizer is
\begin{equation}
    \widetilde f(\boldsymbol{\theta})
    =
    f(\boldsymbol{\theta})+\xi,\qquad
    \xi\sim\mathcal{N}(0,\sigma^2).
\end{equation}
The low-noise setting uses $S_{\rm eff}=8192$ and
$\sigma=0.005524$ in normalized objective units. The high-noise setting uses
$S_{\rm eff}=128$ and $\sigma=0.044194$. These are effective Gaussian
measurement-noise levels rather than an explicit Pauli-term shot-allocation
model.

The optimizer receives only $\widetilde f$. The exact objective is evaluated
offline at every queried point and is stored only for diagnostics. To compare
heterogeneous optimizers under one common rule, the evaluation wrapper stores
the complete query history. Two performance quantities are distinguished:
\begin{align}
    \varepsilon_{\rm oracle}
    &=
    \min_{\theta\ {\rm visited}}
    \varepsilon_{\rm exact}(\theta),\\
    \widehat{\theta}_{\rm hist}
    &=
    \arg\min_{\theta\ {\rm visited}}\widetilde f(\theta),\\
    \varepsilon_{\rm selected}
    &=
    \varepsilon_{\rm exact}(\widehat{\theta}_{\rm hist}).
\end{align}
The oracle metric measures whether the search visited a good exact point and is
not available to the optimizer. The selected metric evaluates a common
single-sample, best-noisy-history rule. It is not necessarily the native return
object of an optimizer: for population methods it can differ from the best
member of the final population, a population mean, or a package-specific
terminal solution. No final re-evaluation or resampling is used in the primary reported
selected metric. Consequently, $\varepsilon_{\rm selected}$ includes both
optimization error and error from selecting the minimum among many noisy
observations. Both noisy studies use the same ten optimizers, 25 runs per
condition, and 10\,000/30\,000 FE ceilings as the deterministic experiment.

As a submission-stage robustness check, we additionally test a retrospective
fixed-budget verification proxy using the stored noisy histories. For a protocol
$(K,R)$, search is truncated after $B-KR$ evaluations, the $K$ stored query
records with the lowest noisy values are retained, and each is assigned $R$
fresh independent observations from the same Gaussian model. The candidate with
the smallest remeasurement mean is selected. We test
$(K,R)\in\{(5,10),(5,20),(10,20)\}$; the $K=10,R=20$ protocol reserves only 200
of 30\,000 evaluations. Because the original histories do not store every
queried parameter vector, the retained set is record-level and can contain
repeat occurrences of the same candidate. The analysis is therefore a
retrospective test of the statistical remeasurement principle, not an exact
reconstruction of a distinct-candidate hardware protocol.

\subsection{Three-instance family robustness extension}
\label{sec:family-methods}

The full ten-method benchmark intentionally uses one median-rugged $N=12$, $p=3$ representative from each family. To test how much of the reported ordering is representative-specific, we add a targeted robustness extension using the two other $N=12$ candidates that were already generated and screened for each family. No new instance-selection rule is introduced: the same Hamiltonian generators, deterministic instance seeds, QAOA simulator, bounds, FE accounting, and low/high-noise definitions are reused. The extension therefore contains three physical instances per family---the original selected representative plus the two unused screen candidates.

To limit computation while retaining the main competing search mechanisms, five methods are rerun on the two added instances: MS-BFGS, MS-CMA-ES, iL-SHADE, L-SRTDE, and jSO-lite. Each added instance uses 10 runs at $30\,000$ FEs under exact, low-noise, and high-noise observations; the original representative contributes its existing 25 runs. Family summaries are formed in two stages: first take the median across optimizer restarts within each physical instance, then give the three instance medians equal weight. Thus the 25-run representative does not outweigh either 10-run added instance. For optimizer comparisons we use one common variational reference per physical instance, updated by the best exact energy visited anywhere in the combined exact/noisy data, so the family-level optimization error does not include differences in finite-depth variational gap.

All retained runs in this five-method extension consume the full 30\,000-FE allowance, so differences in the extension are not attributable to method-specific early termination. With only three physical instances per family, however, this is a robustness check rather than a powered family-level inference study. Pairwise instance-level sign tests are therefore reported only descriptively in Appendix~\ref{app:family}; optimizer restarts are not treated as independent family replicates.

\subsection{Structure-aware tree-search extension}
\label{sec:mcts-methods}

A supplementary experiment reuses exactly the four selected $N=12$, $p=3$ representatives and the same independent $10\,000$- and $30\,000$-FE budgets. The purpose is not to enlarge the primary ten-optimizer ranking, but to test whether QAOA-specific cross-depth structure changes the optimization picture. We use an Agirre-style discretization with branching factor $b=30$, UCT exploration constant $C=\sqrt{2}$, and exponential reward scale $\nu=1/2$ \cite{Agirre2025MCTS}. The extension contains seven exact-objective controls:
\begin{itemize}
    \item \textbf{MCTS-max}: unrestricted sequential MCTS over the full discretized angle box;
    \item \textbf{SSR-MCTS}: iterative MCTS with the depth-$p$ search region restricted using the solution found at depth $p-1$;
    \item \textbf{SSR-random}: the same restricted grids sampled uniformly, isolating the contribution of the MCTS tree policy;
    \item \textbf{MCTS-BFGS}: MCTS used to propose promising leaves, followed by continuous BFGS refinement;
    \item \textbf{Interp-1BFGS}: one BFGS trajectory initialized from a circular interpolation of the previous-depth solution;
    \item \textbf{Interp-BFGS}: the same interpolated center followed by jittered local restarts; and
    \item \textbf{RandomCenter-BFGS}: the same jittered local-search mechanism around a random center, controlling for the effect of neighborhood multistart alone.
\end{itemize}
For iterative methods, all shallower-depth evaluations are charged to the same end-to-end FE budget. The stage allocation is proportional to $1000+800(2p-1)$, matching the depth-dependent evaluation schedule used as the reference MCTS protocol; therefore $B_1+B_2+B_3=B$ for the present $p=3$ benchmark. The $10\,000$ and $30\,000$ cases are run independently. Twenty-five replicates are used per family and budget.

For standalone MCTS, two outcomes are kept distinct. The \emph{oracle best visited} value is the lowest exact objective among all evaluated leaves and is diagnostic only. The primary returned outcome is the exact objective of the strategy selected by the final tree policy. This distinction is important because a tree can visit a good leaf without assigning sufficient value or visits to the prefix leading to it. Under noisy observations, only MCTS-max and SSR-MCTS are repeated at the two noise levels used in the main benchmark, preserving the same oracle-versus-selected distinction. Additional tree diagnostics and the full ablation are placed in Appendix~\ref{app:mcts}.

\section{Results}

\subsection{Landscape and variational limits}

The four selected $N=12$, $p=3$ landscapes differ despite having the same six-dimensional parameterization. Max-Cut has the fewest distinct BFGS-quench endpoint levels ($K_E=6$), while EA, diluted three-spin, and SK have 18, 23, and 16. The near-best catchment fraction is also larger for Max-Cut (0.292) than for EA and diluted three-spin (0.042 each). SK has the largest random-line minimum count and active Hessian condition number. These diagnostics are summarized in Table~\ref{tab:screen}.

\begin{table}[!htbp]
\centering
\caption{Landscape diagnostics for the selected $N=12$, $p=3$ representatives.
$K_E$ is the number of distinct endpoint energy levels from 24 BFGS quenches.}
\label{tab:screen}
\begin{tabular}{lrrrrr}
\toprule
Model & $K_E$ & $p_{0.01}$ & Line minima & $\kappa_{\rm act}$ & med. $g_{\rm RMS}$ \\
\midrule
Max-Cut 3-regular & 6 & 0.292 & 9.5 & 32.5 & 0.223 \\
2D Edwards--Anderson & 18 & 0.042 & 10.0 & 27.5 & 0.172 \\
Diluted 3-spin glass & 23 & 0.042 & 11.5 & 25.9 & 0.045 \\
Sherrington--Kirkpatrick & 16 & 0.167 & 16.0 & 75.9 & 0.126 \\
\bottomrule
\end{tabular}
\end{table}

\begin{figure}[!htbp]
\centering
\includegraphics[width=\textwidth]{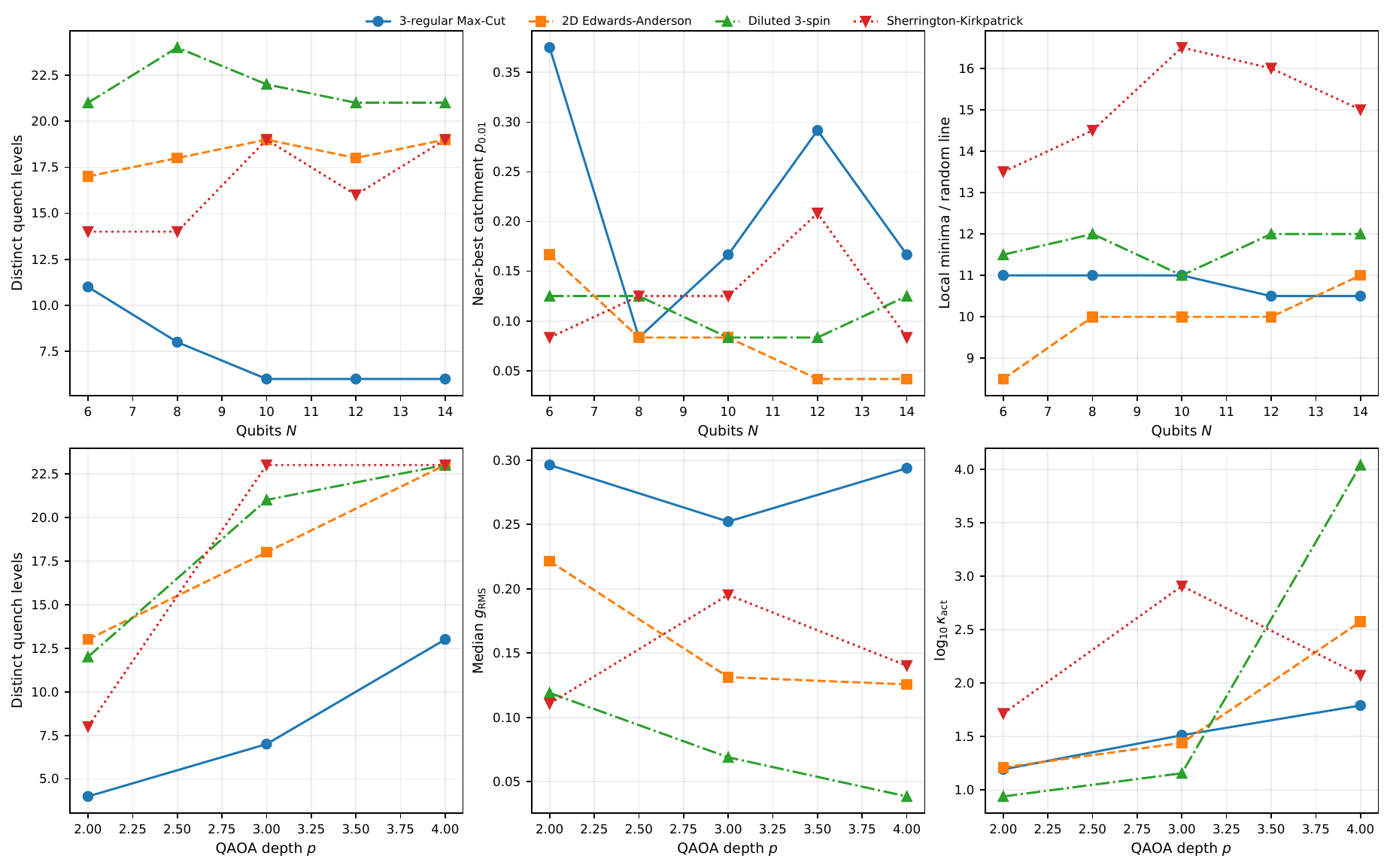}
\caption{Shared-QAOA landscape characterization. Top row: size dependence at
fixed depth $p=3$. Bottom row: depth dependence of quench-level multiplicity,
random-point gradient scale, and active Hessian condition number. Line styles
and markers distinguish the four Hamiltonian families.}
\label{fig:landscape}
\end{figure}

Figure~\ref{fig:landscape} gives the size and depth controls. The selected $N=12$ cost Hamiltonians have 18, 46, 81, and 22 classical one-spin-flip local minima for Max-Cut, EA, diluted three-spin, and SK, respectively. No selected family shows a collapse of the sampled gradient scale to zero.

The $p=3$ variational reference remains above the physical ground state in every model. The normalized variational gaps are 0.1045 for Max-Cut, 0.0976 for EA, 0.1979 for diluted three-spin, and 0.1554 for SK (Table~\ref{tab:refs}).

\begin{table}[!htbp]
\centering
\caption{Variational references for the selected $N=12$, $p=3$ instances.}
\label{tab:refs}
\begin{tabular}{lrrr}
\toprule
Model & $E_0$ & $E_{\rm ref}$ & $\varepsilon_{\rm var}$ \\
\midrule
Max-Cut 3-reg. & -14 & -10.656868 & 0.1045 \\
2D EA & -13 & -10.462143 & 0.0976 \\
Diluted 3-spin & -14 & -8.460157 & 0.1979 \\
SK & -24 & -16.540590 & 0.1554 \\
\bottomrule
\end{tabular}
\end{table}

\subsection{Optimizer benchmark}

\paragraph{Exact objective.}
With exact objectives, restart-based local methods are strongest overall. At 10\,000 FEs, MS-BFGS or MS-CMA-ES reaches numerical reference accuracy on Max-Cut, while jSO-lite is best by median error on EA and L-SRTDE is best on diluted three-spin. MS-BFGS is best on SK. The full medians are in Table~\ref{tab:10kfull}.

\begin{table}[!htbp]
\centering
\caption{Median normalized optimization error over 25 runs at 10\,000 FEs.}
\label{tab:10kfull}
\resizebox{\textwidth}{!}{
\begin{tabular}{lrrrrrrrrrr}
\toprule
Model & BFGS & MS-BFGS & CMA-ES & MS-CMA-ES & SPSA & iSOMA & SciPy-DE & iL-SHADE & L-SRTDE & jSO-lite \\
\midrule
Max-Cut 3-reg. & $0.0175$ & $<5\times10^{-15}$ & $0.0111$ & $5.47\times10^{-14}$ & $0.0419$ & $0.0303$ & $0.0181$ & $0.0128$ & $0.0426$ & $0.0026$ \\
2D EA & $0.0780$ & $0.0209$ & $0.0723$ & $0.0221$ & $0.0900$ & $0.0417$ & $0.0255$ & $0.0221$ & $0.0269$ & $0.0026$ \\
Diluted 3-spin & $0.0694$ & $0.0176$ & $0.0400$ & $0.0176$ & $0.1105$ & $0.0401$ & $0.0197$ & $0.0176$ & $0.0142$ & $0.0176$ \\
SK & $0.0252$ & $5.18\times10^{-15}$ & $0.0126$ & $0.0106$ & $0.0568$ & $0.0411$ & $0.0242$ & $0.0121$ & $0.0361$ & $0.0109$ \\
\bottomrule
\end{tabular}}
\end{table}

\begin{figure}[!htbp]
\centering
\includegraphics[width=\textwidth]{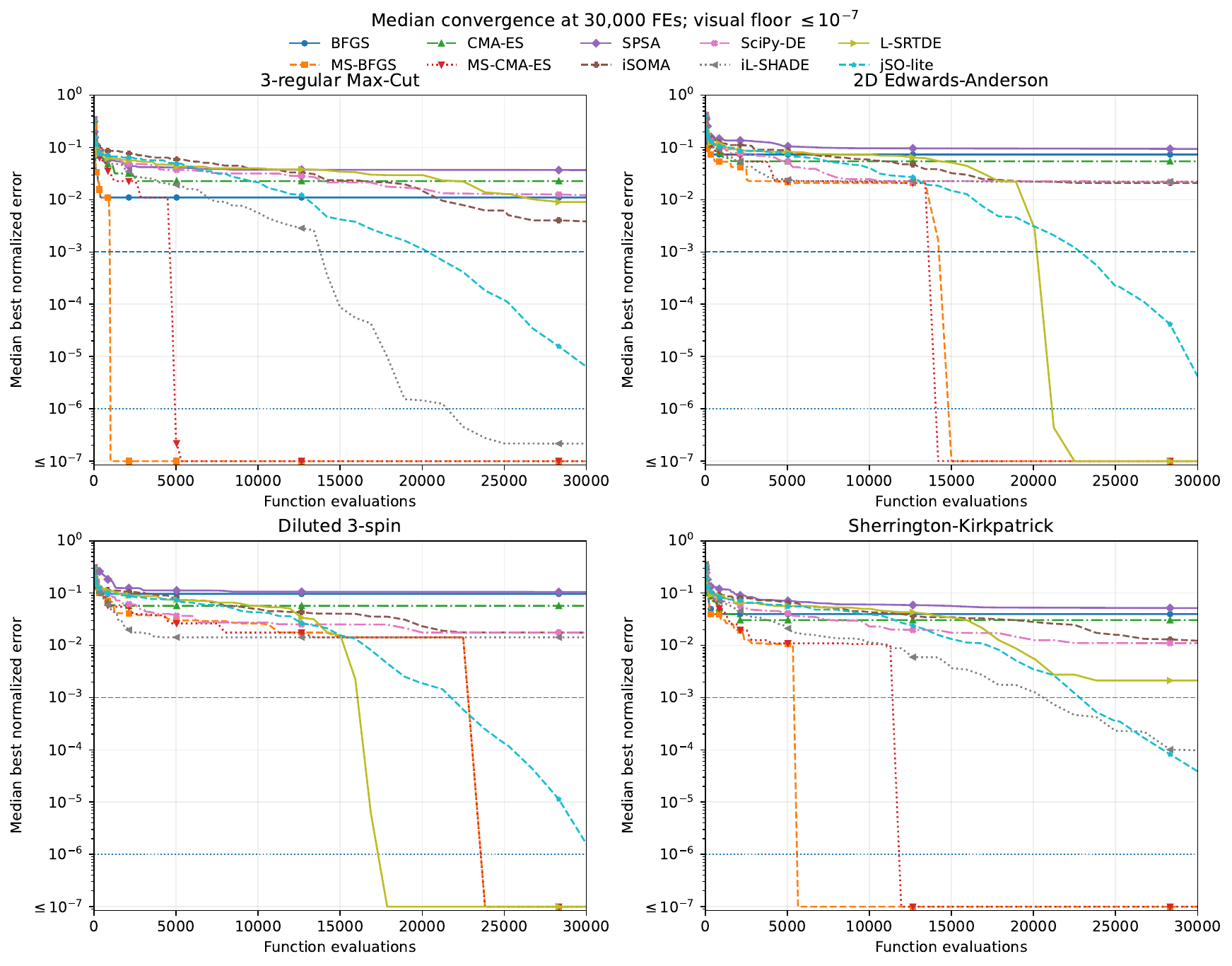}
\caption{Median best-so-far normalized optimization error over 25 runs at
30\,000 FEs. The visual floor is $10^{-7}$; values below this threshold are
not visually resolved. Horizontal reference levels indicate the $10^{-3}$
and $10^{-6}$ accuracy targets.}
\label{fig:convergence}
\end{figure}

At 30\,000 FEs, MS-BFGS and MS-CMA-ES reach numerical reference accuracy on all four models. L-SRTDE also reaches this level on EA and diluted three-spin. jSO-lite and iL-SHADE remain competitive but usually converge more gradually (Table~\ref{tab:30kfull} and Figure~\ref{fig:convergence}).

\begin{table}[!htbp]
\centering
\caption{Median normalized optimization error over 25 runs at 30\,000 FEs.}
\label{tab:30kfull}
\resizebox{\textwidth}{!}{
\begin{tabular}{lrrrrrrrrrr}
\toprule
Model & BFGS & MS-BFGS & CMA-ES & MS-CMA-ES & SPSA & iSOMA & SciPy-DE & iL-SHADE & L-SRTDE & jSO-lite \\
\midrule
Max-Cut 3-reg. & $0.0111$ & $<5\times10^{-15}$ & $0.0228$ & $2.22\times10^{-14}$ & $0.0370$ & $0.0039$ & $0.0122$ & $2.16\times10^{-07}$ & $0.0091$ & $6.40\times10^{-06}$ \\
2D EA & $0.0734$ & $<5\times10^{-15}$ & $0.0543$ & $1.27\times10^{-13}$ & $0.0938$ & $0.0209$ & $0.0221$ & $0.0221$ & $<5\times10^{-15}$ & $4.20\times10^{-06}$ \\
Diluted 3-spin & $0.0964$ & $<5\times10^{-15}$ & $0.0568$ & $6.18\times10^{-14}$ & $0.1049$ & $0.0176$ & $0.0176$ & $0.0142$ & $<5\times10^{-15}$ & $1.59\times10^{-06}$ \\
SK & $0.0396$ & $<5\times10^{-15}$ & $0.0303$ & $4.78\times10^{-14}$ & $0.0515$ & $0.0123$ & $0.0110$ & $9.82\times10^{-05}$ & $0.0021$ & $3.88\times10^{-05}$ \\
\bottomrule
\end{tabular}}
\end{table}

The target-hit probabilities show the same distinction between basin access and high-precision refinement. On Max-Cut, MS-BFGS and MS-CMA-ES reach both $10^{-3}$ and $10^{-6}$ in all 30\,000-FE runs. On EA and diluted three-spin, jSO-lite frequently reaches $10^{-3}$ but reaches $10^{-6}$ less often. Endpoint distributions and exact-objective statistical tests are in Appendices~\ref{app:endpoints} and \ref{app:statistics}.

\begin{figure}[!htbp]
\centering
\includegraphics[width=\textwidth]{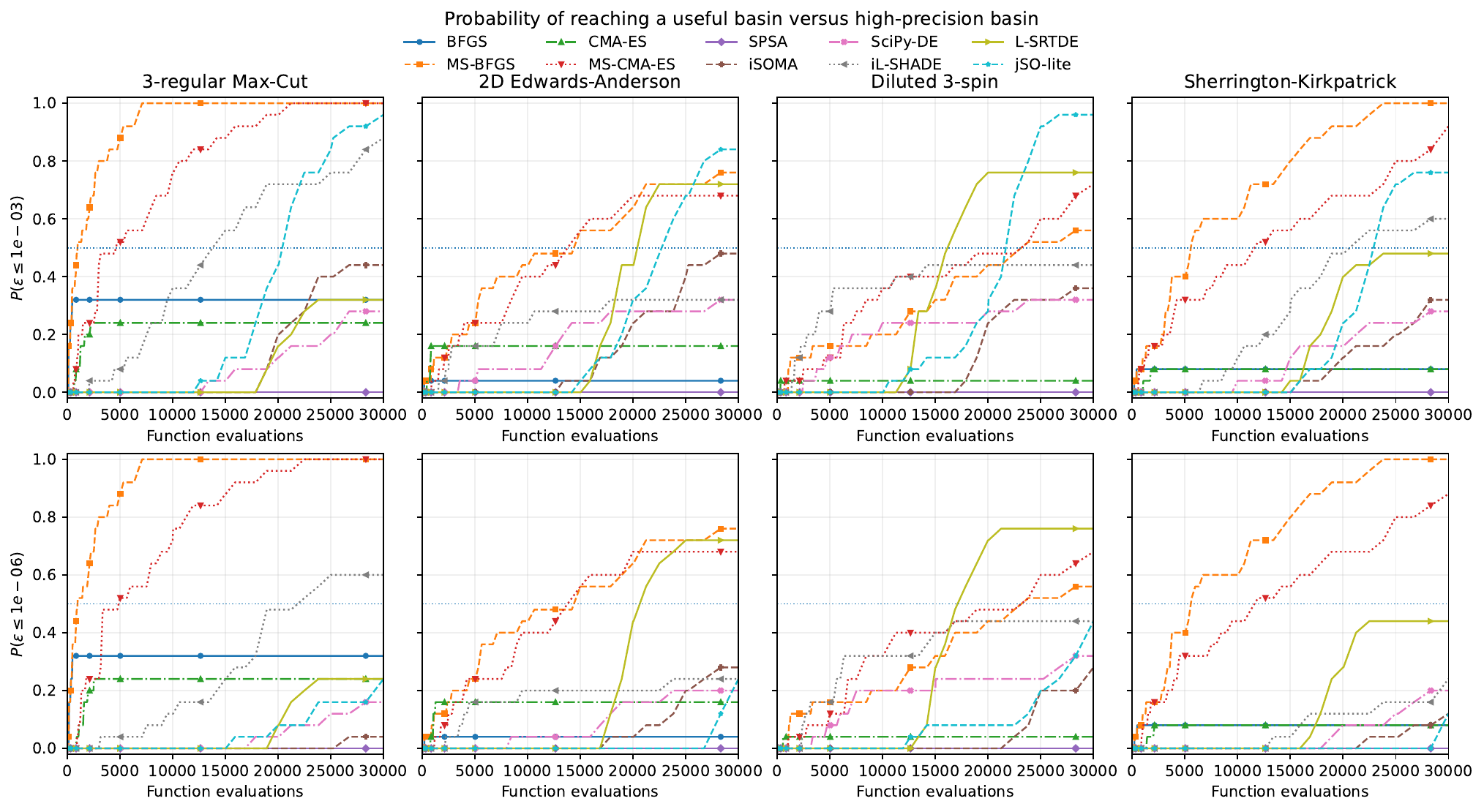}
\caption{Probability of reaching normalized optimization-error targets as a
function of FE count at the 30\,000-FE budget. Top row:
$\varepsilon_{\rm opt}\leq10^{-3}$; bottom row:
$\varepsilon_{\rm opt}\leq10^{-6}$. Probabilities are computed from the
unclipped numerical errors.}
\label{fig:success}
\end{figure}

\FloatBarrier
\paragraph{Low observation noise.}
On the originally selected representatives, at $S_{\rm eff}=8192$ ($\sigma=0.005524$), the exact-objective ordering changes immediately. At 10\,000 FEs, iL-SHADE has the lowest median oracle error on Max-Cut ($0.0141$), diluted three-spin ($0.0177$), and SK ($0.0136$), while jSO-lite is lowest on EA ($0.0206$). By 30\,000 FEs, the longer-horizon adaptive DE methods separate more clearly: jSO-lite is lowest on Max-Cut ($0.00220$) and SK ($0.00470$), whereas L-SRTDE is lowest on EA ($9.59\times10^{-4}$) and diluted three-spin ($0.00128$). The common selected-point rule changes the model-level winners: L-SRTDE has the lowest 30\,000-FE selected median on all four models.

The landscape diagnostics help organize these differences. Max-Cut has only six quench endpoint levels and the largest near-best catchment fraction, so exact multistart local search covers its important basins efficiently. EA and diluted three-spin have many more endpoint levels and a near-best catchment fraction of only $0.042$; on these two models L-SRTDE continues to gain from 10\,000 to 30\,000 FEs and reaches median oracle errors close to $10^{-3}$. SK has the largest line-minimum count and active Hessian condition number; under low noise, jSO-lite and L-SRTDE are the two lowest 30\,000-FE oracle medians. These associations are consistent with population search being useful when narrow or multiple basins must be covered, but the experiment does not isolate landscape geometry as a causal variable.

\begin{figure}[!htbp]
\centering
\includegraphics[width=\textwidth]{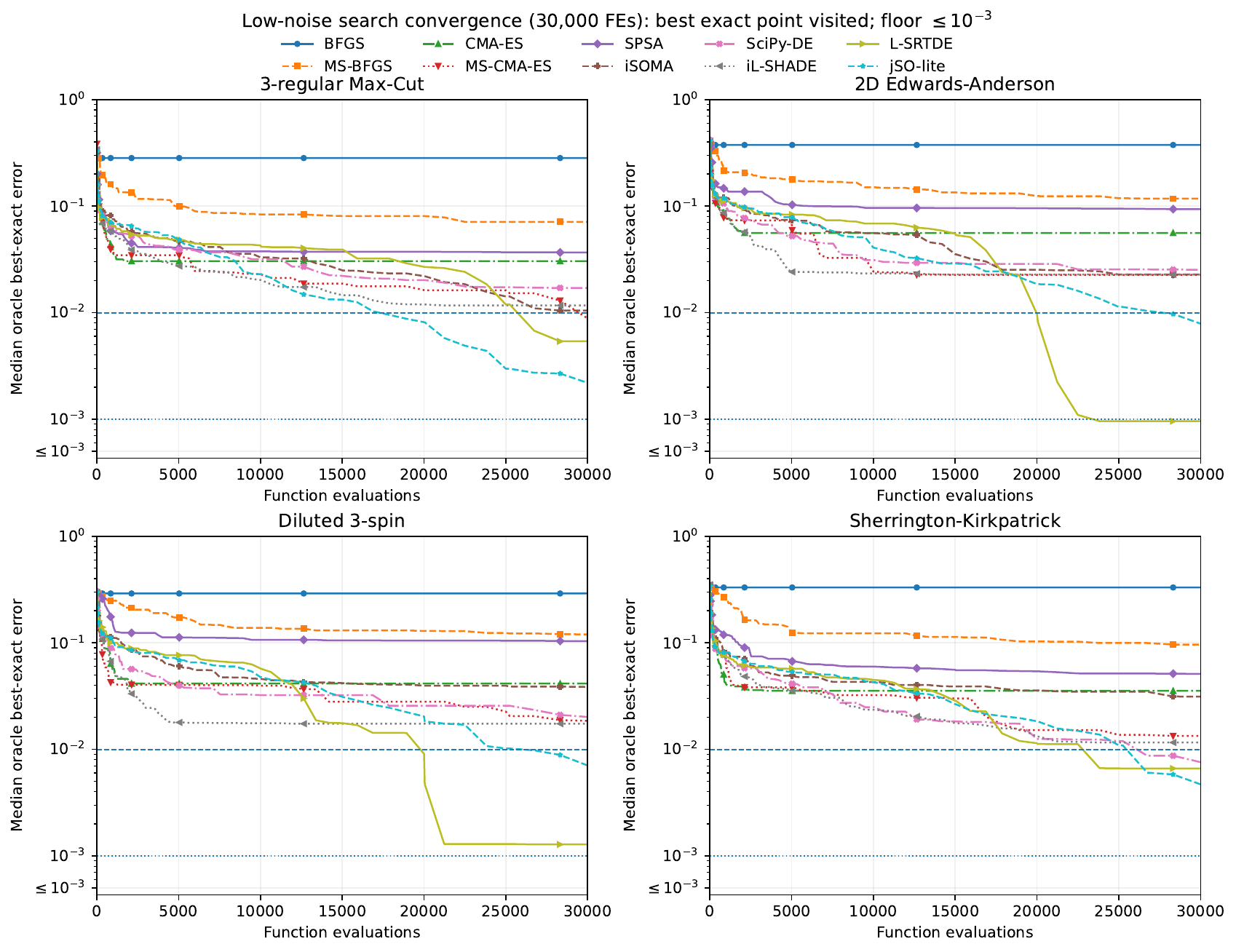}
\caption{Low-noise optimizer convergence at 30\,000 FEs. Curves show the median exact error of the best point visited, evaluated offline; the optimizer receives only noisy values. Values below $10^{-4}$ are visually clipped.}
\label{fig:noiseconv}
\end{figure}

The model-level medians and convergence curves show that the ranking changes materially with landscape and budget. Because the optimizer panel mixes deliberately paired restart controls with otherwise heterogeneous stochastic methods, the revised manuscript treats these broad cross-method differences descriptively rather than relying on an all-method independence test. Figure~\ref{fig:noiseconv} shows the convergence dynamics.

\begin{figure}[!htbp]
\centering
\includegraphics[width=\textwidth]{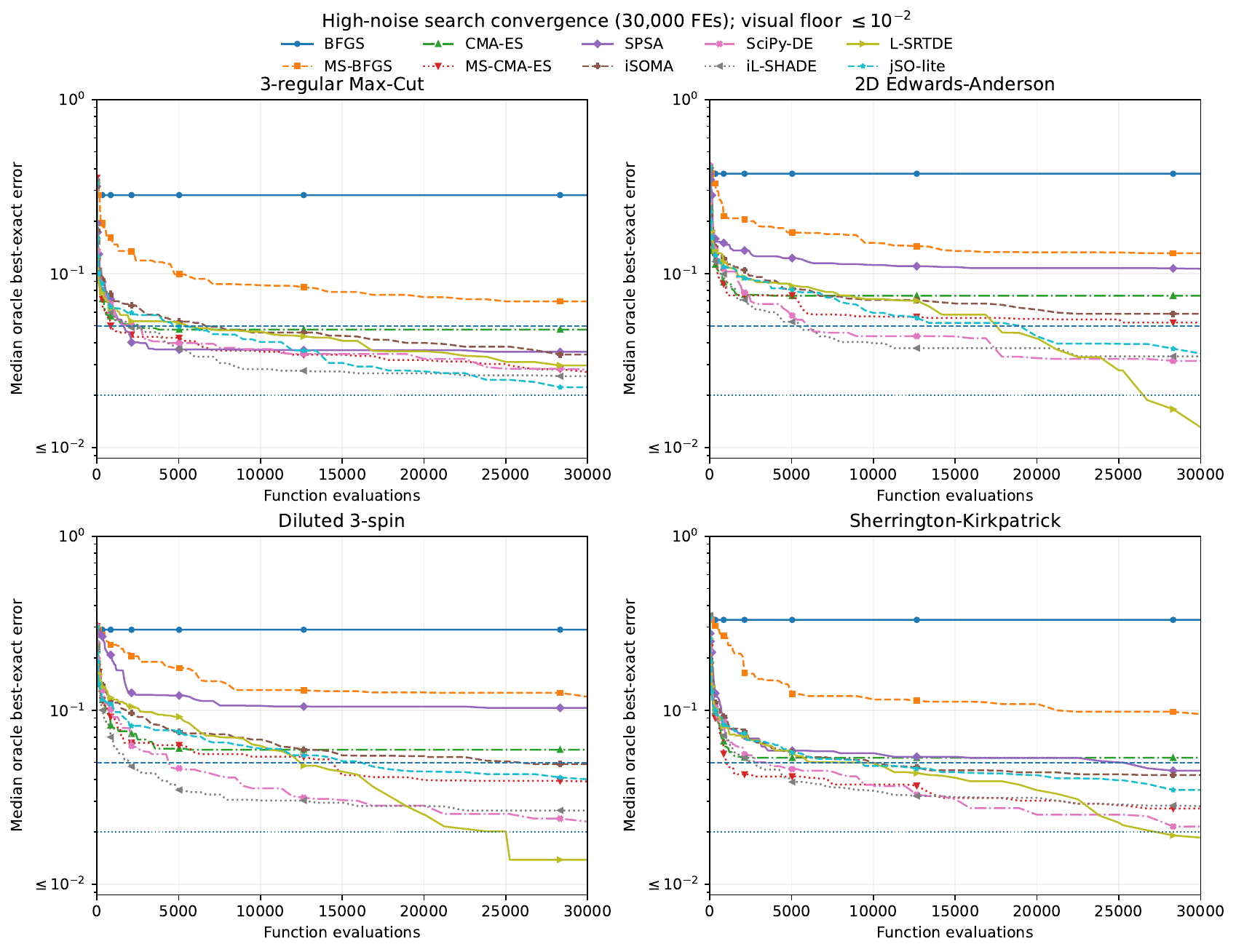}
\caption{High-noise search convergence at 30\,000 FEs. Curves show the median exact error of the best point visited; the optimizer receives only noisy values.}
\label{fig:highnoiseconv}
\end{figure}

\FloatBarrier
\paragraph{High observation noise.}
On the originally selected representatives, at $S_{\rm eff}=128$ ($\sigma=0.044194$), iL-SHADE has the numerically lowest 10\,000-FE oracle median on all four models; on Max-Cut it is effectively tied with MS-CMA-ES ($0.031428$ versus $0.031429$). The ordering changes with additional budget. At 30\,000 FEs, jSO-lite is lowest on Max-Cut ($0.0222$), while L-SRTDE is lowest on EA ($0.0131$), diluted three-spin ($0.0138$), and SK ($0.0185$). Thus iL-SHADE is strongest as a short-budget high-noise search method, whereas L-SRTDE benefits more from the longer run on the three more structured spin-glass landscapes. The high-noise convergence curves are shown in Figure~\ref{fig:highnoiseconv}.

The high-noise selected-point metric shows a second effect that is not visible from oracle convergence. MS-CMA-ES has the best aggregate selected rank across the eight model--budget conditions because it remains near the front at both budgets. At 30\,000 FEs, however, the lowest selected median is SPSA on Max-Cut ($0.0437$) and L-SRTDE on EA ($0.0319$), diluted three-spin ($0.0246$), and SK ($0.0407$). SPSA does not show the same behavior on the more rugged models: its corresponding selected medians are $0.1259$, $0.1124$, and $0.0685$. This isolated Max-Cut result is therefore not evidence for general SPSA superiority under high noise.

\begin{figure}[!htbp]
\centering
\includegraphics[width=\textwidth]{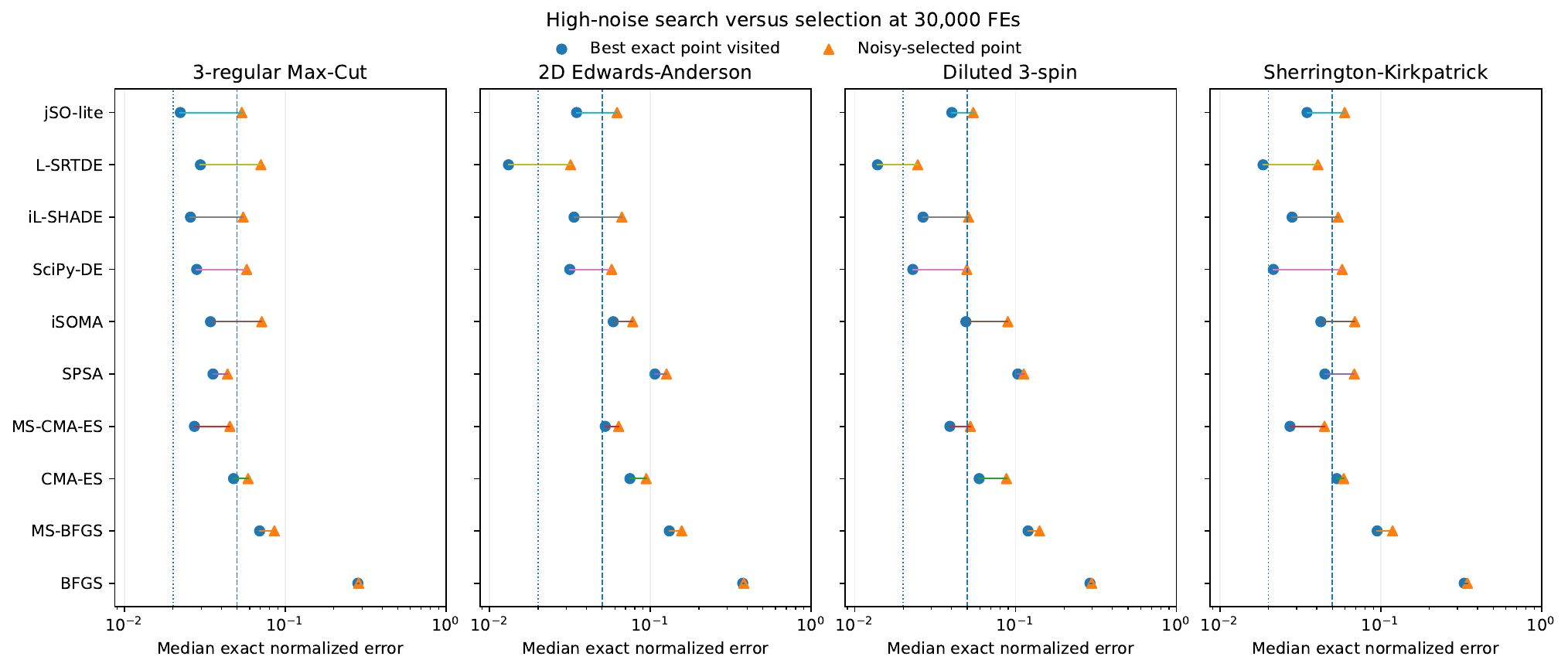}
\caption{High-noise search versus selection at 30\,000 FEs. Circles show the median exact error of the best point visited; triangles show the median exact error of the point selected from noisy observations.}
\label{fig:highnoisegap}
\end{figure}

Selection error becomes comparable to search error in this regime. For the 30\,000-FE oracle leaders, the median exact error changes from $0.0222$ to $0.0536$ for jSO-lite on Max-Cut, from $0.0131$ to $0.0319$ for L-SRTDE on EA, from $0.0138$ to $0.0246$ on diluted three-spin, and from $0.0185$ to $0.0407$ on SK. Figure~\ref{fig:highnoisegap} displays this separation directly. The corresponding distributional differences are reported descriptively through medians, ranks, and endpoint distributions; formal paired inference is reserved for the deliberately matched restart controls.

\begin{table}[!htbp]
\centering
\small
\caption{Noisy leaders on the originally selected representatives at 30\,000 FEs. Values are median normalized optimization errors over 25 runs. Oracle uses the best exact point visited; selected uses the exact error of the common best-noisy-history choice. The three-instance robustness extension is summarized separately in Table~\ref{tab:family-extension-ranks}.}
\label{tab:noiseleaders}
\setlength{\tabcolsep}{4pt}
\begin{tabular}{lrrrr}
\toprule
Model & Low oracle & Low selected & High oracle & High selected \\
\midrule
Max-Cut & jSO-lite $\;0.00220$ & L-SRTDE $\;0.00718$ & jSO-lite $\;0.0222$ & SPSA $\;0.0437$ \\
2D EA & L-SRTDE $\;0.000959$ & L-SRTDE $\;0.00259$ & L-SRTDE $\;0.0131$ & L-SRTDE $\;0.0319$ \\
Diluted 3-spin & L-SRTDE $\;0.00128$ & L-SRTDE $\;0.00379$ & L-SRTDE $\;0.0138$ & L-SRTDE $\;0.0246$ \\
SK & jSO-lite $\;0.00470$ & L-SRTDE $\;0.00792$ & L-SRTDE $\;0.0185$ & L-SRTDE $\;0.0407$ \\
\bottomrule
\end{tabular}
\end{table}

\FloatBarrier

\subsection{Robustness across three instances per family}
\label{sec:family-results}

The targeted three-instance extension preserves the broad regime change but weakens several named-optimizer conclusions drawn from a single representative. Table~\ref{tab:family-extension-ranks} reports mean ranks over the four equal-instance family medians for the five rerun methods. Under exact objectives, MS-BFGS and MS-CMA-ES remain the two strongest methods (mean family ranks 1.50 and 1.75), confirming that restart-based local/covariance search is not an artifact of the originally selected instances. Under low noise, jSO-lite has the best oracle aggregate rank (1.25), whereas under high noise iL-SHADE has the best oracle aggregate rank (1.50). For the high-noise selected-point metric, L-SRTDE ranks first in all four family summaries among the five rerun methods, giving mean rank 1.00.

\begin{table}[!htbp]
\centering
\caption{Three-instance family robustness at 30\,000 FEs. Entries are mean ranks over the four family-level medians (1 is best) after giving each physical instance equal weight. Oracle columns use the best exact point visited; selected columns use the exact quality of the point chosen from the noisy history. Only the five methods rerun in the family extension are included.}
\label{tab:family-extension-ranks}
\begin{tabular}{lrrrrr}
\toprule
Method & Exact & Low oracle & Low selected & High oracle & High selected \\
\midrule
MS-BFGS & 1.50 & 5.00 & 5.00 & 5.00 & 5.00 \\
MS-CMA-ES & 1.75 & 2.25 & 2.25 & 3.00 & 2.50 \\
iL-SHADE & 4.00 & 2.75 & 2.75 & 1.50 & 2.75 \\
L-SRTDE & 4.75 & 3.75 & 3.50 & 2.50 & 1.00 \\
jSO-lite & 3.00 & 1.25 & 1.50 & 3.00 & 3.75 \\
\bottomrule
\end{tabular}
\end{table}

The oracle--selected split is important for interpreting the last result. Across the four high-noise family summaries, the median selected-minus-oracle gap is about $0.0104$ for L-SRTDE, compared with $0.0138$ for MS-CMA-ES, $0.0228$ for iL-SHADE, and $0.0260$ for jSO-lite. Thus L-SRTDE's selected-point advantage is not the same as having the strongest oracle search: iL-SHADE finds the best points most consistently, while L-SRTDE loses less quality under the common noisy-history selection rule.

The extension also shows why family-level language should be used cautiously. On the original representative, L-SRTDE had the lowest low-noise oracle median on EA and diluted three-spin; after adding the two unused instances, the equal-instance family median is instead lowest for jSO-lite on both families. Likewise, the high-noise oracle leader changes from L-SRTDE to iL-SHADE for the EA and diluted-three-spin family medians. These changes do not overturn the main exact-versus-noisy conclusion: multistart methods dominate the exact regime, while adaptive population methods dominate the noisy regime. They do show that the identity of the best adaptive variant is substantially instance dependent even within one Hamiltonian family.

Because the family extension contains only three physical instances, no pairwise sign test can provide strong family-level evidence; the minimum nonzero two-sided sign-test $p$-value with three unanimous instance wins is 0.25 before multiplicity correction. Detailed equal-instance errors, representative-to-family rank changes, and per-instance distributions are therefore placed in Appendix~\ref{app:family} and interpreted descriptively.

\FloatBarrier

\subsection{Cross-regime summary}

Table~\ref{tab:unified} gives the common rank view for the original full ten-method matrix on the selected representatives across models, budgets, and noise levels. The three-instance robustness extension in Section~\ref{sec:family-results} is deliberately separate because it contains only five methods at $30\,000$ FEs. Under exact evaluations, MS-BFGS ranks first overall (1.500) and MS-CMA-ES second (2.625). Under low noise, jSO-lite ranks first by oracle best-visited error (2.000), followed by L-SRTDE and iL-SHADE (2.625 each). Under high noise, iL-SHADE ranks first by oracle error (2.000), followed by SciPy-DE (2.750) and MS-CMA-ES (3.000).

\begin{table}[!htbp]
\centering
\caption{Unified optimizer ranking across models and noise levels. Each model entry is the mean rank over the 10\,000- and 30\,000-FE budgets; 1 is best. Avg. averages all eight model--budget conditions. Noisy ranks use the exact error of the best point visited (oracle search metric). The exact quality of the point selected from noisy observations is summarized in Appendix~\ref{app:noise}. MC: Max-Cut; EA: Edwards--Anderson; 3S: diluted three-spin; SK: Sherrington--Kirkpatrick.}
\label{tab:unified}
\scriptsize
\setlength{\tabcolsep}{2.4pt}
\resizebox{\textwidth}{!}{
\begin{tabular}{lrrrrr rrrrr rrrrr}
\toprule
& \multicolumn{5}{c}{Exact objective}
& \multicolumn{5}{c}{Low noise}
& \multicolumn{5}{c}{High noise} \\
\cmidrule(lr){2-6}\cmidrule(lr){7-11}\cmidrule(lr){12-16}
Optimizer & MC & EA & 3S & SK & Avg.
& MC & EA & 3S & SK & Avg.
& MC & EA & 3S & SK & Avg. \\
\midrule
BFGS & 6.50 & 9.00 & 9.00 & 8.00 & 8.125 & 10.00 & 10.00 & 10.00 & 10.00 & 10.000 & 10.00 & 10.00 & 10.00 & 10.00 & 10.000 \\
MS-BFGS & 1.00 & 2.00 & 2.00 & 1.00 & \textbf{1.500} & 9.00 & 9.00 & 9.00 & 9.00 & 9.000 & 9.00 & 9.00 & 9.00 & 9.00 & 9.000 \\
CMA-ES & 6.50 & 8.00 & 7.50 & 6.50 & 7.125 & 6.00 & 6.00 & 6.50 & 6.50 & 6.250 & 6.00 & 7.00 & 6.50 & 6.00 & 6.375 \\
MS-CMA-ES & 2.00 & 3.00 & 3.50 & 2.00 & 2.625 & 3.50 & 5.00 & 4.50 & 5.00 & 4.500 & 2.50 & 4.00 & 3.00 & 2.50 & 3.000 \\
SPSA & 9.50 & 10.00 & 10.00 & 10.00 & 9.875 & 8.00 & 8.00 & 8.00 & 8.00 & 8.000 & 6.00 & 8.00 & 8.00 & 7.00 & 7.250 \\
iSOMA & 6.50 & 6.00 & 7.50 & 8.00 & 7.000 & 5.00 & 5.50 & 6.50 & 6.50 & 5.875 & 7.00 & 6.00 & 6.50 & 7.00 & 6.625 \\
SciPy-DE & 7.50 & 6.00 & 6.00 & 6.00 & 6.375 & 4.50 & 4.50 & 4.00 & 3.50 & 4.125 & 3.50 & 2.00 & 3.00 & 2.50 & 2.750 \\
iL-SHADE & 4.00 & 5.00 & 4.00 & 4.00 & 4.250 & 3.00 & 3.00 & 2.00 & 2.50 & 2.625 & 1.50 & 2.00 & 2.00 & 2.50 & \textbf{2.000} \\
L-SRTDE & 8.00 & 3.50 & 1.00 & 6.50 & 4.750 & 4.50 & 2.50 & 1.50 & 2.00 & 2.625 & 6.00 & 3.00 & 2.00 & 3.00 & 3.500 \\
jSO-lite & 3.50 & 2.50 & 4.50 & 3.00 & 3.375 & 1.50 & 1.50 & 3.00 & 2.00 & \textbf{2.000} & 3.50 & 4.00 & 5.00 & 5.50 & 4.500 \\
\bottomrule
\end{tabular}}
\end{table}

\FloatBarrier
The rank transition is systematic. Exact evaluations favor restart-based basin coverage: MS-BFGS ranks first overall and MS-CMA-ES second. Low-noise oracle search shifts the lead to jSO-lite, with iL-SHADE and L-SRTDE close behind. High-noise oracle search favors iL-SHADE, SciPy-DE, MS-CMA-ES, and L-SRTDE, while the selected-point ranking favors MS-CMA-ES. Thus the main distinction is not simply ``local versus global'': restart-based CMA-ES remains competitive under high noise, whereas numerical-gradient BFGS loses its deterministic advantage.

The model columns also show that the same noise level does not produce one optimizer ordering. L-SRTDE is strongest on EA and diluted three-spin at the longer budgets and becomes the high-noise leader on SK, while Max-Cut retains competitive solutions from jSO-lite, MS-CMA-ES, and even SPSA under the selected metric. This mirrors the landscape screen: Max-Cut has broader near-best catchment, whereas the spin-glass models have more endpoint levels, more line minima, or stronger conditioning. The connection is descriptive rather than causal because only one representative instance per family is benchmarked in the full optimizer matrix.

The oracle--selected distinction is quantitatively small at low noise but becomes a separate performance axis at high noise. At 30\,000 FEs, the equal-family median selected-minus-oracle gap is $0.0149\,[0.0111,0.0185]$ for MS-CMA-ES, $0.0160\,[0.0140,0.0230]$ for L-SRTDE, $0.0258\,[0.0216,0.0302]$ for iL-SHADE, and $0.0281\,[0.0206,0.0329]$ for jSO-lite; intervals are restart-bootstrap 95\% intervals with the four benchmark families held fixed. The full ten-method table is reported in Appendix~\ref{app:noise}.

A retrospective fixed-budget verification check supports the practical relevance of this gap. With $K=10$ retained records and $R=20$ fresh measurements per record, only 200 of the 30\,000 evaluations are reserved for verification. Under high noise, this proxy lowers the equal-family median selected error for all ten optimizers. Across methods, the median absolute reduction is $0.0069$, corresponding to a median closure of about 35\% of the pre-verification oracle--selected gap. The low-noise effect is much smaller. Because the archived histories do not contain every queried parameter vector, this is a record-level remeasurement proxy rather than an exact distinct-candidate reconstruction; Appendix~\ref{app:noise} gives the full protocol and results.

Detailed noisy medians, endpoint distributions, target probabilities, uncertainty intervals, and verification results remain in Appendix~\ref{app:noise}.

\FloatBarrier
\subsection{Structure-aware tree search}
\label{sec:mcts-results}

The supplementary structure-aware experiment is kept separate from the primary ten-method ranking because it changes the information available to the optimizer by exploiting solutions at shallower QAOA depths. Reprocessed with the same common variational-reference error used in the main benchmark, the $30\,000$-FE exact results show a clear separation between locating useful regions and returning a good discretized tree strategy. Interp-BFGS and MCTS-BFGS reach the numerical reference on the median condition, RandomCenter-BFGS gives $0.0071$, SSR-random $0.0457$, and Interp-1BFGS $0.0490$. By contrast, the strategies returned by MCTS-max and SSR-MCTS have median condition errors $0.307$ and $0.314$, although their best visited leaves are much better ($0.0526$ and $0.0603$).

Thus, in these low-depth conditions, the useful QAOA-specific information is primarily \emph{where to search}: cross-depth restriction and subsequent continuous basin refinement are more effective than the tested standalone tree-selection rule. Noisy MCTS shows the same search--selection separation and is therefore treated as a mechanistic extension rather than a new competitive baseline. The full ablation, noisy results, and common-reference plots are reported in Appendix~\ref{app:mcts}.

\FloatBarrier
\section{Discussion}

The central result is a regime change rather than a universal optimizer ordering. At fixed $D=6$, the four Hamiltonian families exhibit different basin multiplicity, near-best catchment, one-dimensional ruggedness, and local conditioning. With exact objectives, repeated basin coverage followed by local refinement is highly effective: MS-BFGS and MS-CMA-ES lead the aggregate ranking. Once objective noise is introduced, adaptive population methods become comparatively stronger, while numerical-gradient BFGS loses its deterministic advantage. The landscape diagnostics organize these observations, but they do not establish a causal mapping from a particular geometric descriptor to an optimizer.

The three-instance robustness extension strengthens the regime-level conclusion and weakens overly specific family-to-optimizer claims. Exact optimization remains dominated by the multistart methods, while noisy oracle search remains strongest among adaptive population methods. However, the leading adaptive variant changes across physical instances: the equal-instance summaries favor jSO-lite at low noise and iL-SHADE for high-noise oracle search, whereas L-SRTDE gives the strongest high-noise selected-point ranking among the five rerun methods. The appropriate conclusion is therefore that family identity and noise regime constrain the useful search mechanism, but do not uniquely determine a best optimizer. Three instances per family are sufficient as a robustness check against one atypical representative, not for high-powered family-level inference.

The adaptive-DE comparison should also be read in light of implementation scale. L-SRTDE starts from $NP_{\rm init}=20D=120$ at $D=6$ and then reduces the population, whereas iL-SHADE starts from 24 individuals. Literature on noisy evolutionary optimization provides a plausible mechanism by which larger populations can stabilize noisy comparisons or provide implicit averaging, while also emphasizing that the useful population size is finite and problem dependent \cite{ArnoldBeyer2002,ArnoldBeyer2003,NishidaAkimoto2017,Rakshit2017}. Under a fixed FE ceiling, a larger population also means fewer generations. The present experiment changes population size together with mutation, adaptation, and selection rules, so L-SRTDE's selected-point robustness cannot be attributed causally to population size alone.

The jSO-lite baseline is similarly an explicit algorithmic adaptation rather than a claim to reproduce canonical jSO. A direct paper-faithful jSO control on the same four representatives supports the low-budget motivation for the modification: at 10\,000 exact FEs, the equal-family median oracle error is $0.0068$ for jSO-lite versus $0.0277$ for canonical jSO, with three of the four family-wise paired comparisons remaining significant after Holm correction. At 30\,000 exact FEs both variants reach essentially the same numerical reference, and under low/high observation noise the family-wise paired differences do not survive Holm correction. Thus the streamlined variant matters most in the short exact-budget regime; the manuscript retains the name jSO-lite and does not generalize its advantage beyond the tested configuration.

The oracle--selected gap identifies a distinct practical bottleneck. At high noise, several optimizers visit substantially better points than they select using the common best-noisy-history rule, and the restart-bootstrap intervals show that this gap is not merely a single-run fluctuation. This separation is consistent with broader work showing that finite-shot noise can distort both variational optimization and the identification of a reliable final parameter vector \cite{Scriva2024,Novak2025Reliable}. A retrospective fixed-budget verification proxy provides direct, though model-specific, evidence that final remeasurement can help: reserving 200 of 30\,000 evaluations for $K=10$ records remeasured $R=20$ times lowers the equal-family median selected error for all ten methods under high noise. Across methods, the median absolute reduction is $0.0069$, closing about 35\% of the pre-verification oracle--selected gap. This result should not be overinterpreted: the archived histories do not contain every queried parameter vector, so the top-$K$ set is record-level rather than guaranteed to contain $K$ distinct candidates, and the remeasurements use the same additive Gaussian model rather than hardware-derived shot noise. It nevertheless turns final verification from a purely prospective recommendation into an empirically supported measurement-allocation strategy within the present noise model.

The FE audit clarifies a second protocol point. The benchmark imposes common FE ceilings, not identical realized evaluation counts. The population and multistart methods exhaust essentially the full ceilings, whereas single-start BFGS and CMA-ES often terminate much earlier under their native criteria; for example, at the 30\,000-FE ceiling the median exact-objective usage is 520 FEs for BFGS and 2,040 for CMA-ES. Their comparison with the multistart variants should therefore be understood as a comparison of native single-start termination against restart-based use of the available budget, not as two algorithms forced to consume exactly the same number of evaluations. This is precisely why the paired BFGS/MS-BFGS and CMA-ES/MS-CMA-ES controls are informative.

The structure-aware extension adds a complementary mechanism result without changing the primary ranking. Iterative restriction and cross-depth continuation can identify productive regions, but the tested standalone MCTS policies return substantially worse schedules than the best leaves they visit. Continuous refinement changes this: MCTS-BFGS and interpolation-guided BFGS reach the common variational reference at 30\,000 exact FEs on the median condition, while restriction-only random search also outperforms the returned standalone MCTS schedules. In these conditions, QAOA-specific information is therefore more useful for localization and initialization than for discrete tree selection itself, consistent with the hybrid MCTS/local-refinement perspective of Agirre et al. and with parameter-transfer studies that exploit symmetry or smooth cross-depth structure \cite{Agirre2025MCTS,Lyngfelt2025Transfer,Mele2022}.

Several limitations remain. The full ten-method matrix uses one selected $N=12$ representative per family, while the robustness extension covers three instances for only five methods and one 30\,000-FE budget. The primary optimizer dimension is $D=6$; the $p=4$ landscape control reaches $D=8$ but is not a higher-dimensional optimizer benchmark. Hyperparameters are globally fixed rather than nested-tuned, BFGS uses numerical finite differences, and the noise model is additive Gaussian rather than explicit Pauli-term sampling or hardware noise. Future work should therefore prioritize larger held-out instance ensembles, a focused $D>6$ comparison of the leading methods, nested hyperparameter tuning, analytic-gradient baselines, distinct-candidate verification on archived parameter vectors, and hardware-derived noise. These extensions refine the scope of the present conclusions rather than altering the main empirical observation that optimizer effectiveness depends jointly on landscape, noise, and the rule used to identify a final solution.

\section{Conclusion}

Classical optimization in low-depth QAOA is not governed by a single optimizer ranking. Exact statevector objectives favor restart-based local or covariance-adaptation search, whereas noisy feedback shifts the advantage toward adaptive population methods. The three-instance family extension shows that this regime change is more stable than the identity of any named adaptive-DE winner: jSO-lite leads the low-noise oracle summary, iL-SHADE the high-noise oracle summary, and L-SRTDE the high-noise selected-point summary among the rerun methods. The result is therefore best interpreted at the level of search mechanisms and information use rather than as a universal family-to-optimizer prescription.

The study also separates finding a good point from recognizing it under noise. Bootstrap intervals confirm a substantial high-noise oracle--selected gap for several population methods, and the retrospective fixed-budget remeasurement proxy shows that reserving a small fraction of the measurement budget for final verification can recover part of that loss. The supplementary structure-aware experiment points in the same direction from a different angle: cross-depth restriction and continuous basin refinement are useful, while standalone MCTS selection remains weak in the tested low-depth setting. Together, these results support a landscape- and noise-aware classical loop in which global basin coverage, structural initialization when available, and explicit final-point verification are treated as separate design choices.

\section*{Acknowledgments}
This project has received funding from the Research Council of Lithuania (LMTLT), agreement No. P-ITP-24-9. This research was also supported by research grant SGS No. SP2026/063 of VSB--Technical University of Ostrava, Czech Republic, and by the Ministry of Education, Youth and Sports of the Czech Republic through e-INFRA CZ (ID:90254).

\section*{Data availability}
Code required to reproduce the numerical experiments is available at
\url{https://github.com/VojtechNovak/QAOA-noisy}. Full reproducibility Zenodo package is available at \url{https://doi.org/10.5281/zenodo.22830406}.

\section*{Competing interests}
The authors declare no known competing financial interests or personal relationships that could have influenced the work reported here.

\appendix

\section{Statistical analysis}
\label{app:statistics}

The benchmark contains both independent and deliberately matched comparisons. In particular, BFGS and MS-BFGS share the first initialization within each replicate, and CMA-ES and MS-CMA-ES share both the first initialization and the first CMA-ES random seed. Treating all ten optimizer samples as mutually independent in a single omnibus test would therefore ignore part of the experimental design. The revised inferential analysis is intentionally conservative: broad cross-optimizer results are summarized descriptively by medians, interquartile ranges, ranks, and target-hit probabilities, while the two planned restart comparisons use paired tests.

For each family and FE budget, the restart effect is tested with a two-sided Wilcoxon signed-rank test on the 25 matched runs. Holm correction is applied jointly to the 16 exact-objective restart tests (eight model--budget conditions for BFGS/MS-BFGS and eight for CMA-ES/MS-CMA-ES). Table~\ref{tab:pairedappendix} reports the Holm-adjusted values. The paired conclusions are unchanged: multistart improves the corresponding single-start method in every exact-objective condition.

\begin{table}[!htbp]
\centering
\caption{Holm-adjusted paired Wilcoxon tests for the exact-objective restart controls. All tests use 25 matched runs; adjustment is across the 16 planned restart tests.}
\label{tab:pairedappendix}
\begin{tabular}{lrrr}
\toprule
Model & FE budget & BFGS vs. MS-BFGS & CMA-ES vs. MS-CMA-ES \\
\midrule
Max-Cut & 10,000 & $9.84\times10^{-5}$ & $9.84\times10^{-5}$ \\
Max-Cut & 30,000 & $9.84\times10^{-5}$ & $2.82\times10^{-5}$ \\
EA & 10,000 & $9.84\times10^{-5}$ & $8.17\times10^{-5}$ \\
EA & 30,000 & $9.84\times10^{-5}$ & $1.36\times10^{-5}$ \\
Diluted 3-spin & 10,000 & $9.84\times10^{-5}$ & $9.09\times10^{-5}$ \\
Diluted 3-spin & 30,000 & $9.54\times10^{-7}$ & $1.67\times10^{-6}$ \\
SK & 10,000 & $9.84\times10^{-5}$ & $9.84\times10^{-5}$ \\
SK & 30,000 & $9.54\times10^{-7}$ & $3.87\times10^{-6}$ \\
\bottomrule
\end{tabular}
\end{table}

The same dependence issue applies to the noisy runs because the restart pairs use matched first starts and matched first noise streams. The main text therefore does not use the previously generated all-method Kruskal--Wallis summaries as confirmatory evidence. A future reanalysis can use an explicitly blocked all-method design if a common replicate block is defined for every optimizer; this is separate from the descriptive optimizer ranking reported here.

\section{Endpoint distributions}
\label{app:endpoints}

\begin{figure}[!htbp]
\centering
\includegraphics[width=\textwidth]{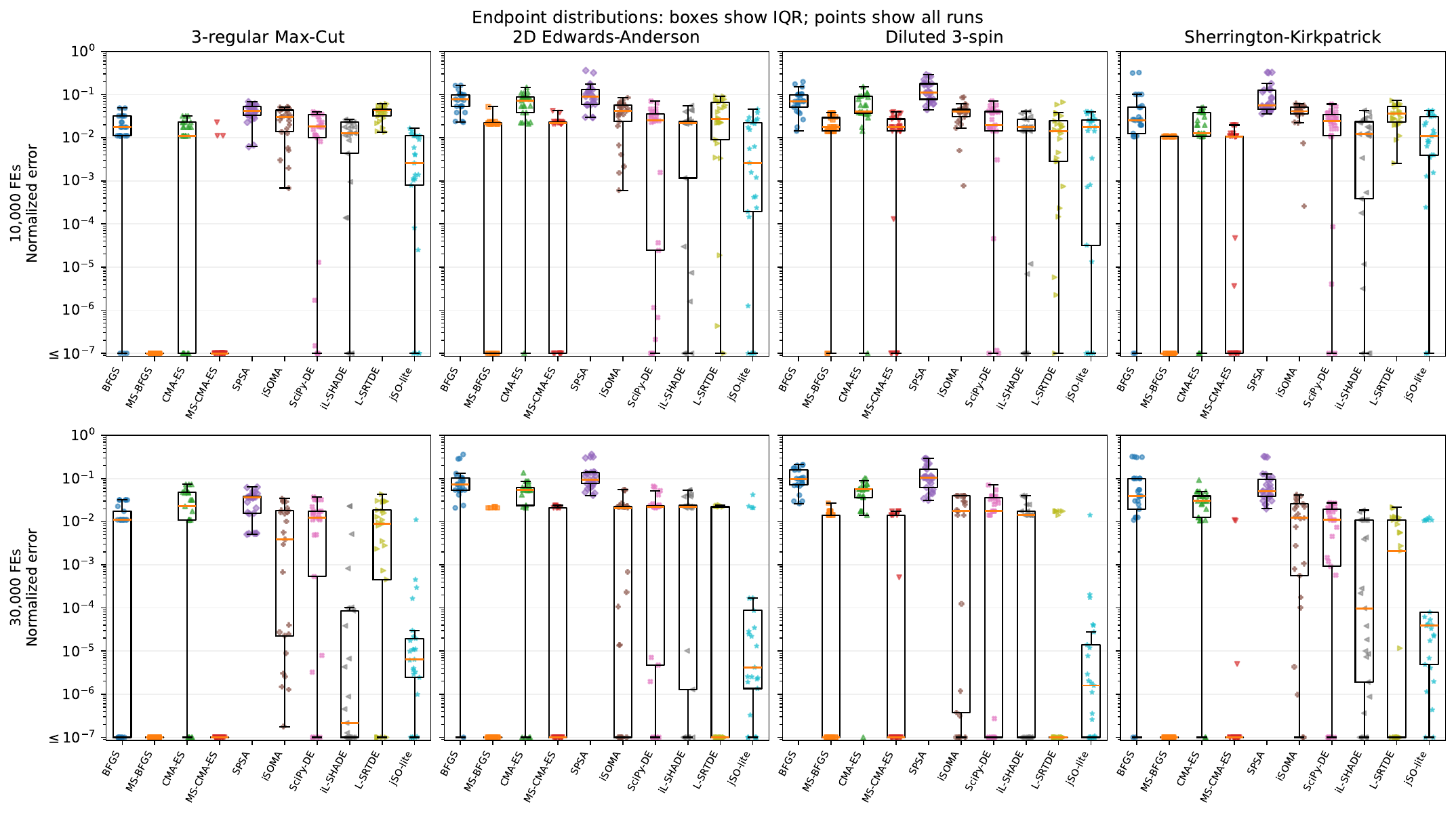}
\caption{Endpoint distributions at 10\,000 and 30\,000 FEs. Boxes show the
interquartile range and median; points show all 25 runs. Errors below
$10^{-7}$ are clipped only for visualization.}
\label{fig:endpointappendix}
\end{figure}

Figure~\ref{fig:endpointappendix} shows all 25 endpoint errors together with
their interquartile ranges. Values below $10^{-7}$ are displayed at the same
plotting floor. The floor therefore denotes high-precision success rather than
additional numerical resolution. EA and diluted three-spin show the strongest
mixture of reference hits and endpoints near $10^{-2}$ for the multistart and
L-SRTDE methods.

\section{Noisy-objective results}
\label{app:noise}

Table~\ref{tab:noise-ranks} compares oracle and selected ranks. The low-noise ordering changes little between the two metrics. At high noise, iL-SHADE is best by oracle rank, while MS-CMA-ES is best by selected rank. Tables~\ref{tab:low10k}--\ref{tab:high30k} give the model-level medians; every noisy cell reports oracle/selected exact error.

\begin{table}[!htbp]
\centering
\caption{Aggregate noisy-objective ranks over eight model--budget conditions. Oracle ranks use the best exact point visited; selected ranks use the exact error of the point chosen by the best-noisy-history rule. Lower is better.}
\label{tab:noise-ranks}
\begin{tabular}{lrrrr}
\toprule
Optimizer & Low oracle & Low selected & High oracle & High selected \\
\midrule
BFGS & 10.000 & 10.000 & 10.000 & 10.000 \\
MS-BFGS & 9.000 & 9.000 & 9.000 & 9.000 \\
CMA-ES & 6.250 & 6.375 & 6.375 & 4.500 \\
MS-CMA-ES & 4.500 & 4.375 & 3.000 & 2.375 \\
SPSA & 8.000 & 8.000 & 7.250 & 5.875 \\
iSOMA & 5.875 & 6.000 & 6.625 & 7.250 \\
SciPy-DE & 4.125 & 4.000 & 2.750 & 3.750 \\
iL-SHADE & 2.625 & 2.500 & 2.000 & 3.125 \\
L-SRTDE & 2.625 & 2.500 & 3.500 & 4.125 \\
jSO-lite & 2.000 & 2.250 & 4.500 & 5.000 \\
\bottomrule
\end{tabular}
\end{table}

\begin{table}[!htbp]
\centering
\caption{Low-noise median normalized error at 10,000 FEs. Each cell is oracle/selected exact error over 25 runs.}
\label{tab:low10k}
\scriptsize
\setlength{\tabcolsep}{2.4pt}
\resizebox{\textwidth}{!}{
\begin{tabular}{lrrrrrrrrrr}
\toprule
Model & BFGS & MS-BFGS & CMA-ES & MS-CMA-ES & SPSA & iSOMA & SciPy-DE & iL-SHADE & L-SRTDE & jSO-lite \\
\midrule
Max-Cut & $0.2929$/$0.2929$ & $0.0829$/$0.0829$ & $0.0302$/$0.0322$ & $0.0236$/$0.0251$ & $0.0421$/$0.0482$ & $0.0319$/$0.0373$ & $0.0222$/$0.0232$ & $0.0141$/$0.0173$ & $0.0374$/$0.0379$ & $0.0166$/$0.0201$ \\
EA & $0.3509$/$0.3565$ & $0.1582$/$0.1653$ & $0.0501$/$0.0513$ & $0.0532$/$0.0545$ & $0.0900$/$0.0935$ & $0.0515$/$0.0566$ & $0.0379$/$0.0379$ & $0.0233$/$0.0250$ & $0.0410$/$0.0424$ & $0.0206$/$0.0234$ \\
Diluted 3-spin & $0.2709$/$0.2819$ & $0.1455$/$0.1458$ & $0.0402$/$0.0433$ & $0.0400$/$0.0409$ & $0.1094$/$0.1111$ & $0.0408$/$0.0428$ & $0.0251$/$0.0310$ & $0.0177$/$0.0192$ & $0.0179$/$0.0208$ & $0.0310$/$0.0326$ \\
SK & $0.2941$/$0.2941$ & $0.1067$/$0.1067$ & $0.0397$/$0.0403$ & $0.0311$/$0.0339$ & $0.0550$/$0.0578$ & $0.0441$/$0.0471$ & $0.0263$/$0.0263$ & $0.0136$/$0.0168$ & $0.0226$/$0.0259$ & $0.0231$/$0.0257$ \\
\bottomrule
\end{tabular}}
\end{table}

\begin{table}[!htbp]
\centering
\caption{Low-noise median normalized error at 30,000 FEs. Each cell is oracle/selected exact error over 25 runs.}
\label{tab:low30k}
\scriptsize
\setlength{\tabcolsep}{2.4pt}
\resizebox{\textwidth}{!}{
\begin{tabular}{lrrrrrrrrrr}
\toprule
Model & BFGS & MS-BFGS & CMA-ES & MS-CMA-ES & SPSA & iSOMA & SciPy-DE & iL-SHADE & L-SRTDE & jSO-lite \\
\midrule
Max-Cut & $0.2825$/$0.2825$ & $0.0712$/$0.0747$ & $0.0304$/$0.0321$ & $0.0088$/$0.0131$ & $0.0368$/$0.0383$ & $0.0105$/$0.0171$ & $0.0171$/$0.0191$ & $0.0117$/$0.0161$ & $0.0054$/$0.0072$ & $0.0022$/$0.0077$ \\
EA & $0.3747$/$0.3754$ & $0.1175$/$0.1175$ & $0.0560$/$0.0592$ & $0.0227$/$0.0231$ & $0.0937$/$0.0968$ & $0.0228$/$0.0253$ & $0.0253$/$0.0306$ & $0.0228$/$0.0248$ & $0.0010$/$0.0026$ & $0.0079$/$0.0108$ \\
Diluted 3-spin & $0.2902$/$0.2902$ & $0.1197$/$0.1197$ & $0.0415$/$0.0423$ & $0.0185$/$0.0194$ & $0.1038$/$0.1063$ & $0.0385$/$0.0416$ & $0.0201$/$0.0237$ & $0.0174$/$0.0186$ & $0.0013$/$0.0038$ & $0.0071$/$0.0107$ \\
SK & $0.3306$/$0.3306$ & $0.0963$/$0.0967$ & $0.0354$/$0.0374$ & $0.0134$/$0.0156$ & $0.0511$/$0.0536$ & $0.0313$/$0.0360$ & $0.0076$/$0.0094$ & $0.0116$/$0.0143$ & $0.0066$/$0.0079$ & $0.0047$/$0.0103$ \\
\bottomrule
\end{tabular}}
\end{table}

\paragraph{Low noise.}
At $S_{\rm eff}=8192$ ($\sigma=0.005524$), jSO-lite has the best aggregate oracle rank. At 30\,000 FEs, the lowest oracle medians are jSO-lite on Max-Cut and L-SRTDE on EA and diluted three-spin; jSO-lite is slightly lower than L-SRTDE on SK by oracle error. For the selected point, L-SRTDE has the lowest 30\,000-FE median on all four models. Figure~\ref{fig:noiseconv} in the main text shows search convergence; Figure~\ref{fig:noisetarget} below gives the corresponding history-selected target probabilities.

\begin{figure}[!htbp]
\centering
\includegraphics[width=\textwidth]{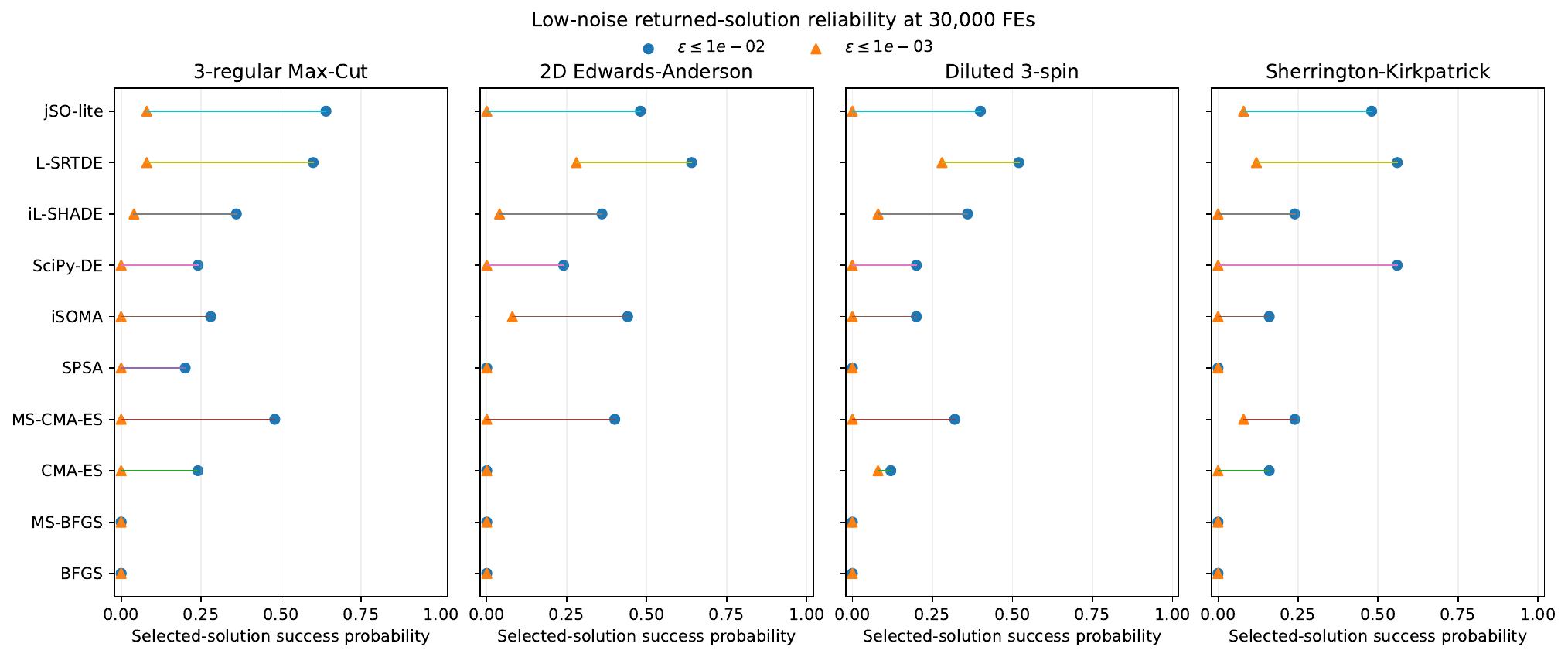}
\caption{Low-noise history-selected solution reliability at 30\,000 FEs. Probabilities refer to the exact error of the point selected by the best-noisy-history rule.}
\label{fig:noisetarget}
\end{figure}

\begin{figure}[!htbp]
\centering
\includegraphics[width=\textwidth]{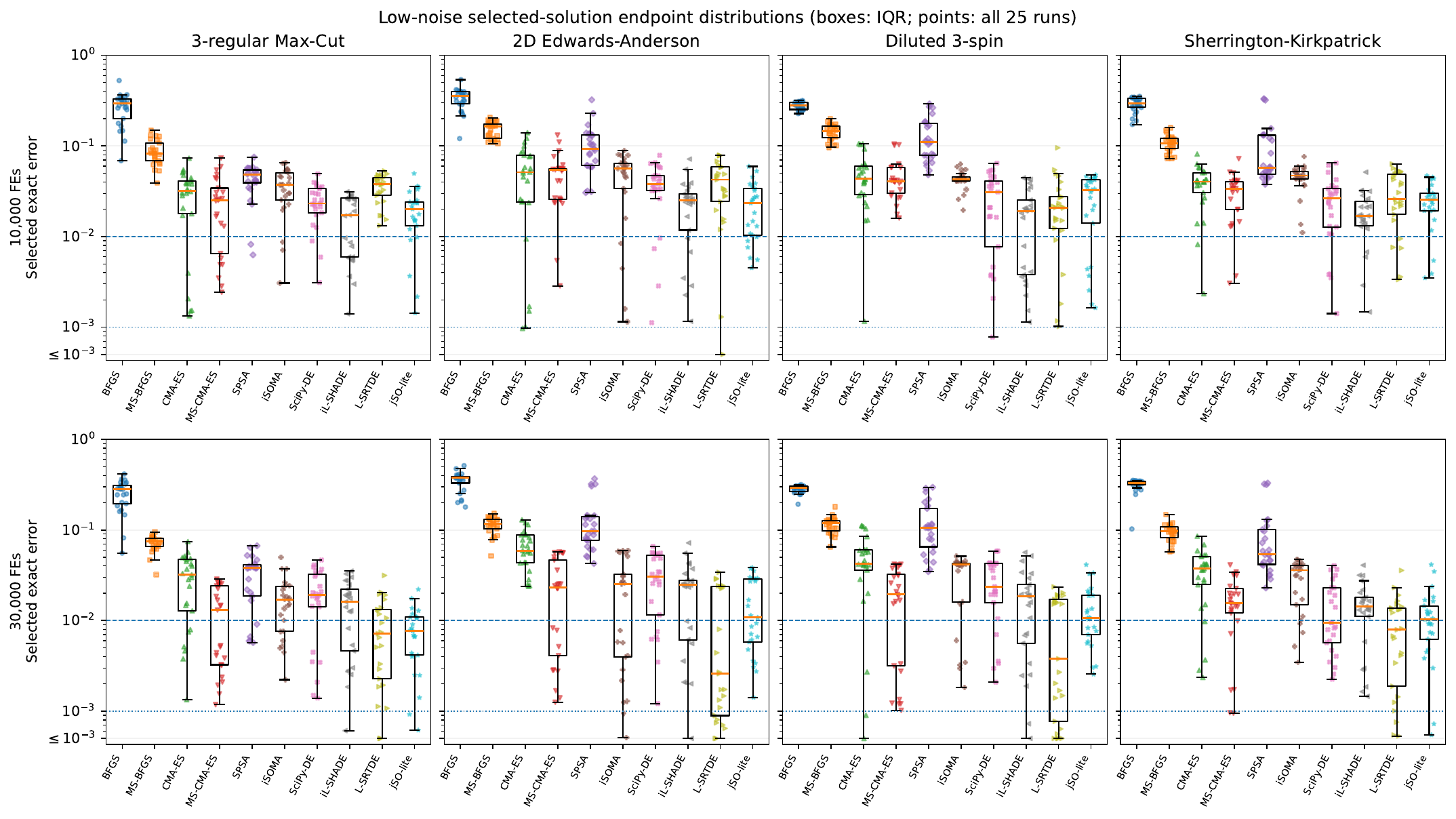}
\caption{Low-noise history-selected endpoint distributions at 10\,000 and 30\,000 FEs. Boxes show the interquartile range and median; points show all 25 runs.}
\label{fig:noiseendappendix}
\end{figure}

\FloatBarrier
\begin{table}[!htbp]
\centering
\caption{High-noise median normalized error at 10,000 FEs. Each cell is oracle/selected exact error over 25 runs.}
\label{tab:high10k}
\scriptsize
\setlength{\tabcolsep}{2.4pt}
\resizebox{\textwidth}{!}{
\begin{tabular}{lrrrrrrrrrr}
\toprule
Model & BFGS & MS-BFGS & CMA-ES & MS-CMA-ES & SPSA & iSOMA & SciPy-DE & iL-SHADE & L-SRTDE & jSO-lite \\
\midrule
Max-Cut & $0.2929$/$0.2993$ & $0.0773$/$0.0888$ & $0.0374$/$0.0520$ & $0.0314$/$0.0511$ & $0.0382$/$0.0483$ & $0.0443$/$0.0772$ & $0.0324$/$0.0581$ & $0.0314$/$0.0576$ & $0.0411$/$0.0740$ & $0.0394$/$0.0754$ \\
EA & $0.3509$/$0.3636$ & $0.1447$/$0.1686$ & $0.0719$/$0.0826$ & $0.0546$/$0.0716$ & $0.1011$/$0.1045$ & $0.0675$/$0.0920$ & $0.0513$/$0.0828$ & $0.0325$/$0.0556$ & $0.0668$/$0.1038$ & $0.0566$/$0.0878$ \\
Diluted 3-spin & $0.2709$/$0.2957$ & $0.1516$/$0.1754$ & $0.0549$/$0.0692$ & $0.0407$/$0.0646$ & $0.1060$/$0.1128$ & $0.0561$/$0.0934$ & $0.0435$/$0.0738$ & $0.0282$/$0.0508$ & $0.0433$/$0.0683$ & $0.0474$/$0.0813$ \\
SK & $0.2941$/$0.3331$ & $0.1067$/$0.1248$ & $0.0408$/$0.0531$ & $0.0354$/$0.0510$ & $0.0485$/$0.0767$ & $0.0499$/$0.1029$ & $0.0385$/$0.0546$ & $0.0336$/$0.0612$ & $0.0461$/$0.0795$ & $0.0479$/$0.0766$ \\
\bottomrule
\end{tabular}}
\end{table}

\begin{table}[!htbp]
\centering
\caption{High-noise median normalized error at 30,000 FEs. Each cell is oracle/selected exact error over 25 runs.}
\label{tab:high30k}
\scriptsize
\setlength{\tabcolsep}{2.4pt}
\resizebox{\textwidth}{!}{
\begin{tabular}{lrrrrrrrrrr}
\toprule
Model & BFGS & MS-BFGS & CMA-ES & MS-CMA-ES & SPSA & iSOMA & SciPy-DE & iL-SHADE & L-SRTDE & jSO-lite \\
\midrule
Max-Cut & $0.2825$/$0.2859$ & $0.0692$/$0.0854$ & $0.0477$/$0.0587$ & $0.0272$/$0.0452$ & $0.0355$/$0.0437$ & $0.0343$/$0.0714$ & $0.0281$/$0.0576$ & $0.0257$/$0.0547$ & $0.0297$/$0.0705$ & $0.0222$/$0.0536$ \\
EA & $0.3747$/$0.3809$ & $0.1310$/$0.1565$ & $0.0745$/$0.0940$ & $0.0523$/$0.0635$ & $0.1066$/$0.1259$ & $0.0587$/$0.0776$ & $0.0315$/$0.0574$ & $0.0335$/$0.0664$ & $0.0131$/$0.0319$ & $0.0347$/$0.0620$ \\
Diluted 3-spin & $0.2902$/$0.2972$ & $0.1196$/$0.1409$ & $0.0594$/$0.0877$ & $0.0390$/$0.0523$ & $0.1032$/$0.1124$ & $0.0491$/$0.0895$ & $0.0229$/$0.0497$ & $0.0266$/$0.0510$ & $0.0138$/$0.0246$ & $0.0401$/$0.0545$ \\
SK & $0.3306$/$0.3460$ & $0.0951$/$0.1185$ & $0.0533$/$0.0589$ & $0.0273$/$0.0447$ & $0.0449$/$0.0685$ & $0.0424$/$0.0690$ & $0.0215$/$0.0576$ & $0.0281$/$0.0543$ & $0.0185$/$0.0407$ & $0.0348$/$0.0597$ \\
\bottomrule
\end{tabular}}
\end{table}

\paragraph{High noise.}
At $S_{\rm eff}=128$ ($\sigma=0.044194$), iL-SHADE has the best aggregate oracle rank. At 10\,000 FEs it has the lowest oracle median on all four models. At 30\,000 FEs, jSO-lite has the lowest oracle median on Max-Cut and L-SRTDE on EA, diluted three-spin, and SK. The selected metric changes the aggregate ordering: MS-CMA-ES ranks first overall, while the lowest 30\,000-FE selected median is SPSA on Max-Cut and L-SRTDE on the other three models. Figures~\ref{fig:highnoiseconv} and \ref{fig:highnoisegap} in the main text show the high-noise convergence and search--selection gap.

\begin{figure}[!htbp]
\centering
\includegraphics[width=\textwidth]{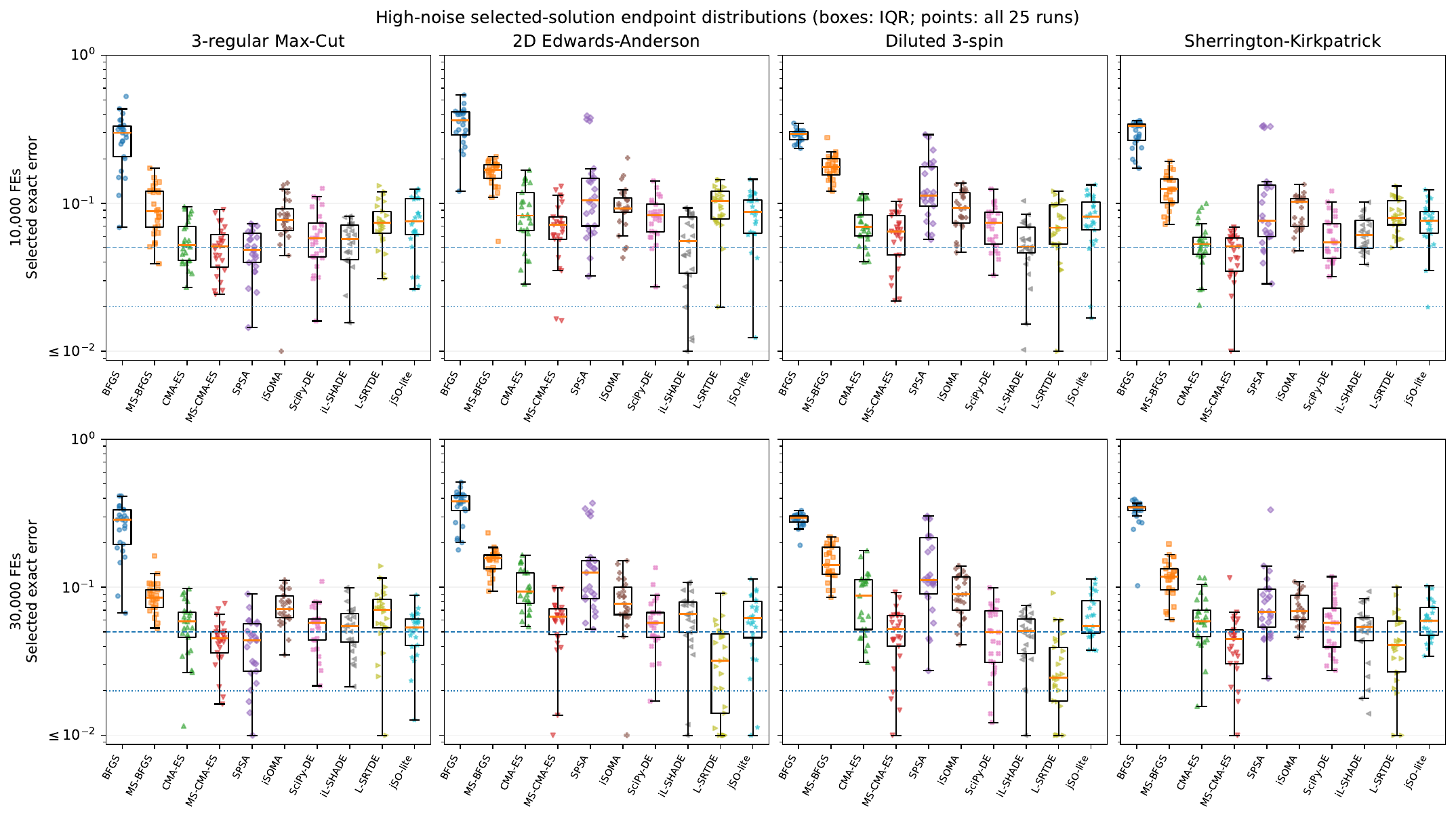}
\caption{High-noise history-selected endpoint distributions at 10\,000 and 30\,000 FEs. Boxes show the interquartile range and median; points show all 25 runs.}
\label{fig:highnoiseendappendix}
\end{figure}

\FloatBarrier

\subsection*{Search--selection uncertainty and retrospective verification}

The selected-minus-oracle gap is summarized with restart-bootstrap intervals in Table~\ref{tab:selection-gap-bootstrap}. The intervals hold the four benchmark families fixed and resample optimizer restarts within each family; they therefore quantify uncertainty of the reported benchmark summary rather than population-level variation over possible Hamiltonian families.

\begin{table}[!htbp]
\centering
\scriptsize
\caption{Selected-minus-oracle optimization-error gap at 30\,000 FEs. Values are the median of the four family-specific medians with 95\% restart-bootstrap intervals; the four benchmark families are held fixed during resampling.}
\label{tab:selection-gap-bootstrap}
\begin{tabular}{lrr}
\toprule
Optimizer & Low noise & High noise\\
\midrule
BFGS & 0.0000 [0.0000,0.0000] & 0.0000 [0.0000,0.0033] \\
MS-BFGS & 0.0000 [0.0000,0.0000] & 0.0150 [0.0076,0.0268] \\
CMA-ES & 0.0013 [0.0011,0.0015] & 0.0107 [0.0091,0.0137] \\
MS-CMA-ES & 0.0012 [0.0010,0.0015] & 0.0149 [0.0111,0.0185] \\
SPSA & 0.0020 [0.0017,0.0028] & 0.0098 [0.0083,0.0124] \\
iSOMA & 0.0023 [0.0016,0.0027] & 0.0299 [0.0233,0.0398] \\
SciPy-DE & 0.0023 [0.0018,0.0028] & 0.0261 [0.0189,0.0313] \\
iL-SHADE & 0.0021 [0.0018,0.0024] & 0.0258 [0.0216,0.0302] \\
L-SRTDE & 0.0013 [0.0009,0.0015] & 0.0160 [0.0140,0.0230] \\
jSO-lite & 0.0040 [0.0031,0.0046] & 0.0281 [0.0206,0.0329] \\
\bottomrule
\end{tabular}
\end{table}

The fixed-budget remeasurement proxy is reported in Table~\ref{tab:verification-proxy} for the most informative tested protocol, $K=10$ and $R=20$. The other tested allocations, $5\times10$ and $5\times20$, give the same qualitative high-noise direction. At 30\,000 FEs the $10\times20$ protocol improves the high-noise equal-family selected median for every optimizer, although the magnitude varies substantially by method. This analysis uses fresh synthetic draws from the same Gaussian model and stored query records; it is not a reconstruction of a hardware measurement campaign or of ten guaranteed-distinct parameter vectors.

\begin{table}[!htbp]
\centering
\small
\caption{Retrospective high-noise final-verification proxy at a fixed 30\,000-FE budget. Search is truncated at 29\,800 FEs and the ten best stored noisy query records are each remeasured 20 times. Values are medians over the four family medians. Because every historical parameter vector was not archived, the top-ten set is record-level and need not contain ten distinct candidates.}
\label{tab:verification-proxy}
\begin{tabular}{lrrr}
\toprule
Optimizer & Best noisy history & Verified proxy & Change\\
\midrule
L-SRTDE & 0.0363 & 0.0304 & -0.0059 \\
MS-CMA-ES & 0.0488 & 0.0442 & -0.0045 \\
iL-SHADE & 0.0545 & 0.0439 & -0.0106 \\
jSO-lite & 0.0571 & 0.0503 & -0.0068 \\
SciPy-DE & 0.0575 & 0.0390 & -0.0185 \\
CMA-ES & 0.0733 & 0.0709 & -0.0024 \\
iSOMA & 0.0745 & 0.0630 & -0.0115 \\
SPSA & 0.0897 & 0.0857 & -0.0040 \\
MS-BFGS & 0.1297 & 0.1120 & -0.0177 \\
BFGS & 0.3216 & 0.3146 & -0.0070 \\
\bottomrule
\end{tabular}
\end{table}

\FloatBarrier

\section{Three-instance family robustness extension}
\label{app:family}

The family robustness extension reuses the two unselected $N=12$, $p=3$ candidates from the original three-candidate screen in each of the four Hamiltonian families. The added instances therefore come from the same generators and deterministic seed convention as the selected representative; no post-hoc hard/easy instance search is performed. Five methods are rerun on each added instance for 10 replicates at $30\,000$ FEs under exact, low-noise, and high-noise observations. The existing 25 replicates for the selected representative are retained, but aggregation is performed at the physical-instance level so that each of the three instances has equal weight.

For this appendix we recompute one common variational reference per physical instance from the best valid reference or exact energy visited anywhere in the combined data. The reported family-level optimizer metric is then the median, across the three physical instances, of each instance's median normalized optimization error relative to that common reference. This differs deliberately from the physical-ground-error plots generated during the raw extension: physical error contains the instance-dependent variational gap and is useful for deployment quality, whereas the common-reference error isolates the classical optimizer comparison.

\begin{table}[!htbp]
\centering
\scriptsize
\caption{Equal-instance family medians of common-reference normalized optimization error at 30\,000 FEs. Each value is the median of the three physical-instance medians.}
\label{tab:family-extension-medians}
\resizebox{\textwidth}{!}{%
\begin{tabular}{lllrrrrr}
\toprule
Regime & Metric & Family & MS-BFGS & MS-CMA-ES & iL-SHADE & L-SRTDE & jSO-lite \\
\midrule
Exact & oracle & Max-Cut & 1.027e-15 & 3.528e-14 & 1.578e-06 & 0.01568 & 6.396e-06 \\
Exact & oracle & EA & 2.186e-15 & 1.848e-13 & 0.02213 & 0.003518 & 1.823e-05 \\
Exact & oracle & 3-spin & 0.009009 & 1.787e-13 & 0.01421 & 0.02384 & 1.592e-06 \\
Exact & oracle & SK & 2.998e-15 & 4.781e-14 & 0.0001157 & 0.01948 & 3.337e-05 \\
Low noise & oracle & Max-Cut & 0.07117 & 0.001606 & 0.0117 & 0.005384 & 0.002197 \\
Low noise & oracle & EA & 0.1175 & 0.02267 & 0.02279 & 0.03419 & 0.005678 \\
Low noise & oracle & 3-spin & 0.1157 & 0.02202 & 0.01741 & 0.03645 & 0.006136 \\
Low noise & oracle & SK & 0.09632 & 0.01336 & 0.0116 & 0.02117 & 0.008362 \\
Low noise & selected & Max-Cut & 0.07471 & 0.003627 & 0.01382 & 0.007177 & 0.007691 \\
Low noise & selected & EA & 0.1175 & 0.0231 & 0.02477 & 0.03657 & 0.0108 \\
Low noise & selected & 3-spin & 0.1157 & 0.02288 & 0.01882 & 0.03752 & 0.008734 \\
Low noise & selected & SK & 0.09671 & 0.01558 & 0.01427 & 0.02217 & 0.01234 \\
High noise & oracle & Max-Cut & 0.06922 & 0.02721 & 0.02153 & 0.02833 & 0.02223 \\
High noise & oracle & EA & 0.124 & 0.04666 & 0.03349 & 0.0421 & 0.0341 \\
High noise & oracle & 3-spin & 0.1196 & 0.03805 & 0.02657 & 0.03736 & 0.03936 \\
High noise & oracle & SK & 0.09908 & 0.02728 & 0.02812 & 0.02513 & 0.03733 \\
High noise & selected & Max-Cut & 0.08543 & 0.03906 & 0.0547 & 0.03683 & 0.05365 \\
High noise & selected & EA & 0.1517 & 0.05714 & 0.05421 & 0.04645 & 0.06199 \\
High noise & selected & 3-spin & 0.1301 & 0.0523 & 0.051 & 0.04817 & 0.05652 \\
High noise & selected & SK & 0.1264 & 0.04479 & 0.05435 & 0.0407 & 0.05915 \\
\bottomrule
\end{tabular}}
\end{table}

\begin{figure}[!htbp]
\centering
\begin{minipage}{0.49\textwidth}
\centering
\includegraphics[width=\linewidth]{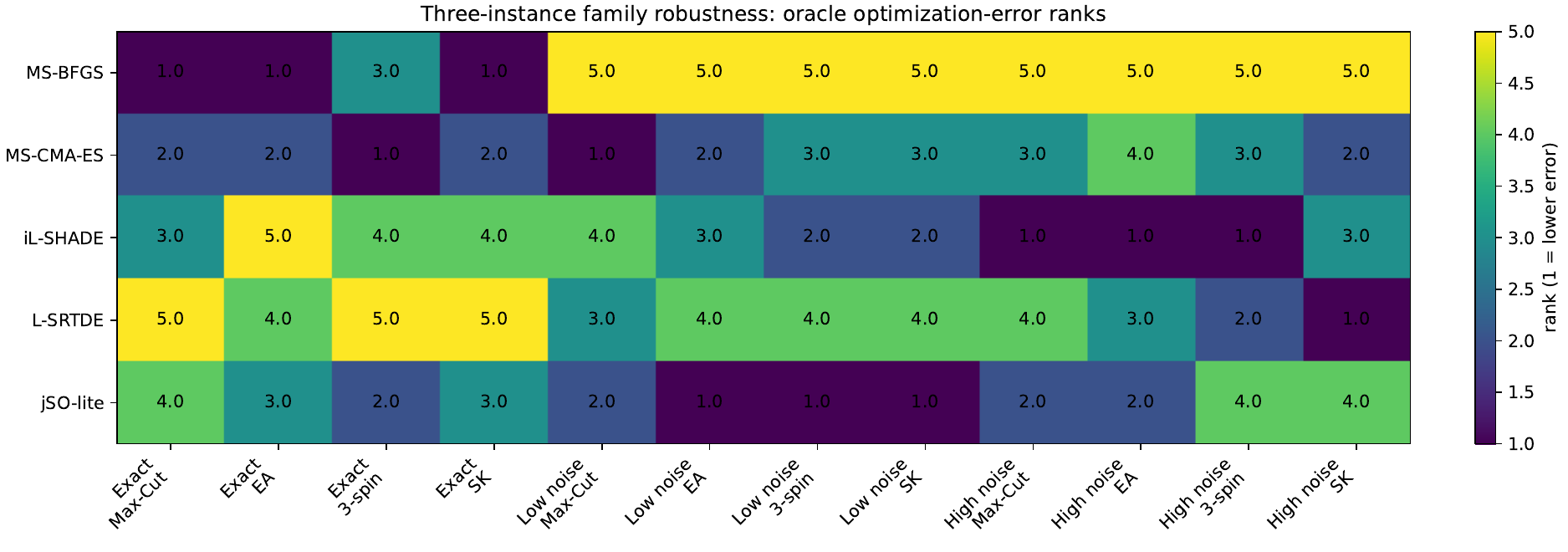}\\[-1mm]
{\small (a) Oracle best-visited rank}
\end{minipage}\hfill
\begin{minipage}{0.49\textwidth}
\centering
\includegraphics[width=\linewidth]{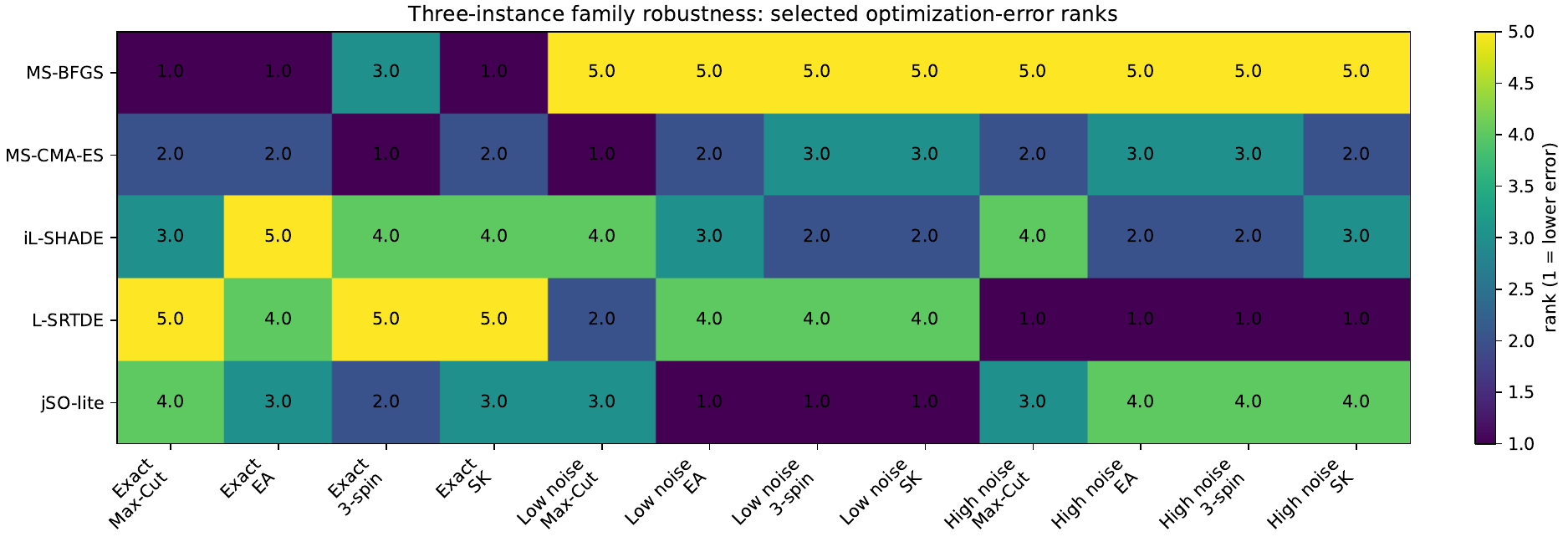}\\[-1mm]
{\small (b) Selected-point rank}
\end{minipage}
\caption{Equal-instance optimizer ranks in the three-instance family extension. Each cell ranks the five rerun methods using the family median of the three physical-instance medians (1 is best). Panel (a) uses the best exact point visited; panel (b) uses the exact quality of the point selected from the noisy history. Exact, low-noise, and high-noise regimes are shown for Max-Cut, EA, diluted three-spin, and SK.}
\label{fig:family-rank-heatmaps}
\end{figure}

Figure~\ref{fig:family-rank-heatmaps} summarizes the equal-instance ranks for oracle search and returned-point quality. The exact row shows the stability of restart-based methods. In the low-noise row, jSO-lite is the strongest aggregate oracle method in the five-method subset, while in the high-noise row iL-SHADE is strongest by oracle search and L-SRTDE by final selected point. The latter difference is another manifestation of the search-versus-identification split discussed in the main text.

\begin{figure}[!htbp]
\centering
\includegraphics[width=0.72\textwidth]{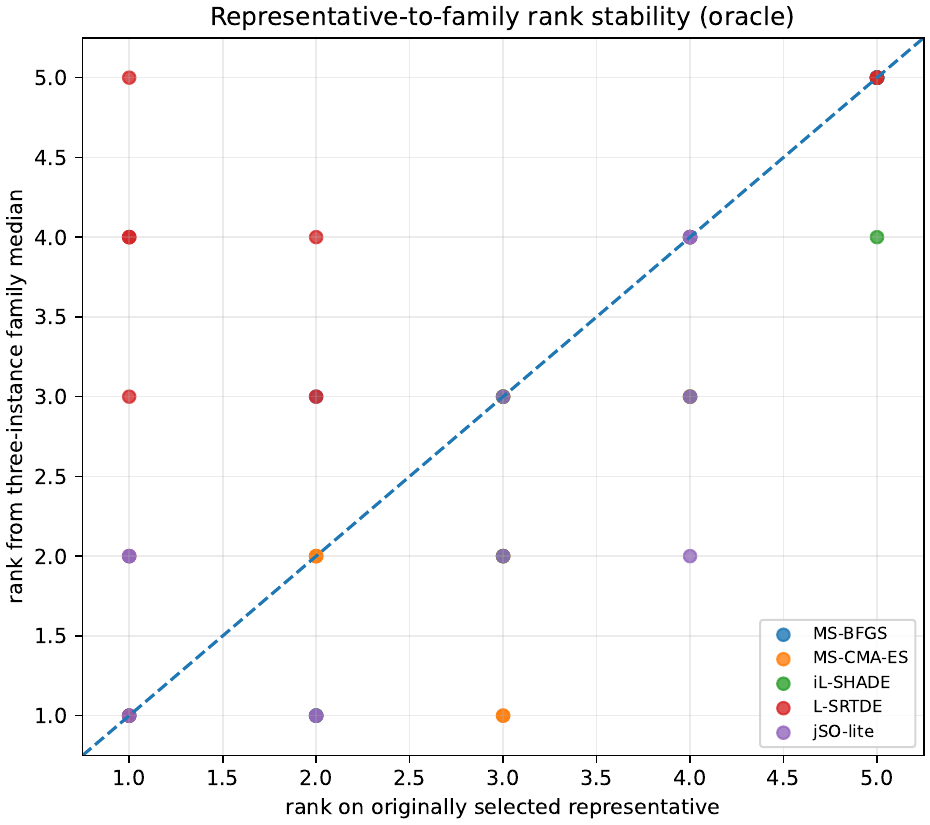}
\caption{Optimizer rank on the originally selected representative versus rank after equal weighting of all three physical instances, pooled over family/noise conditions for the five rerun methods. Points away from the diagonal indicate representative-specific rank changes.}
\label{fig:family-rank-shift}
\end{figure}

The original selected representative does not always preserve the within-family optimizer ordering. This is most visible for EA and diluted three-spin: the selected representative favors L-SRTDE under low-noise oracle search, whereas the three-instance family median favors jSO-lite; at high noise the family-level oracle ordering moves toward iL-SHADE. Figure~\ref{fig:family-rank-shift} shows representative rank against three-instance family rank for the oracle metric. Deviations from the diagonal are precisely the variation that the extension was designed to expose.

The instance count remains too small for strong pairwise family-level inference. With three paired physical instances, even three unanimous wins give a two-sided sign-test $p=0.25$ before Holm correction. We therefore use the extension to assess robustness of qualitative regime conclusions and to expose within-family rank variation, not to certify a universal optimizer ordering for each Hamiltonian family.

\section{Structure-aware MCTS extension}
\label{app:mcts}

This appendix reports the full structure-aware ablation supporting Section~\ref{sec:mcts-results}. All conditions reuse the same selected $N=12$, $p=3$ instances as the primary benchmark, use 25 independent runs, and enforce the same end-to-end FE ceilings. Iterative SSR and interpolation methods pay for the shallower $p=1$ and $p=2$ stages from the stated ceiling; no precomputed continuation schedule is supplied for free. For consistency with the main benchmark, all absolute MCTS values below are re-expressed relative to the common per-condition variational reference rather than to the physical ground state.

\begin{figure}[!htbp]
\centering
\includegraphics[width=0.92\textwidth]{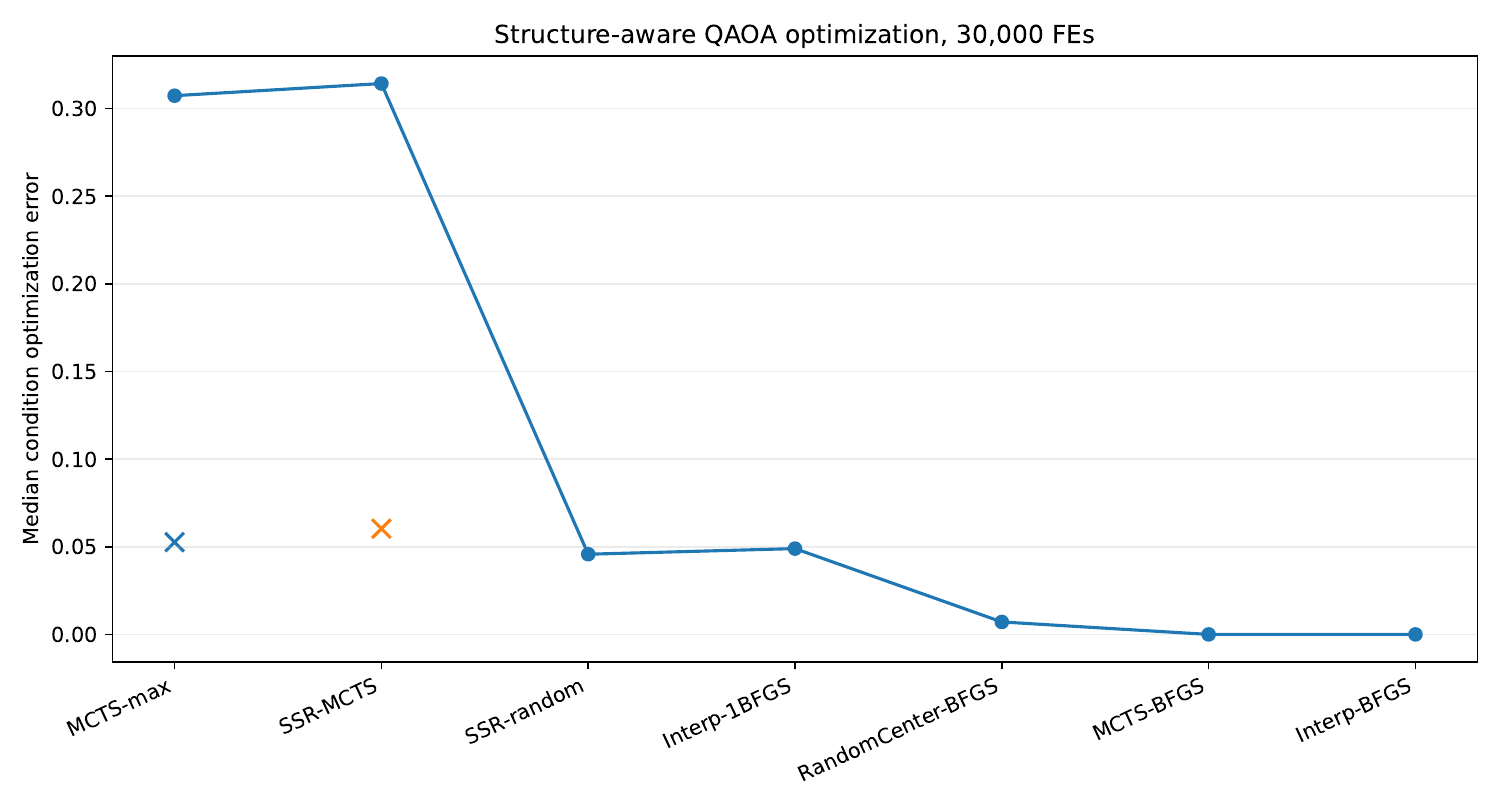}
\caption{Exact-objective structure-aware extension at 30\,000 FEs using common-reference optimization error. Circles show the returned/selected result. Crosses mark the oracle best-visited value for the two standalone MCTS methods, for which selected and oracle outcomes differ substantially.}
\label{fig:mcts-exact-app}
\end{figure}

At 30\,000 FEs, the median across the four condition medians is numerically at the common reference for Interp-BFGS and MCTS-BFGS, $0.0071$ for RandomCenter-BFGS, $0.0457$ for SSR-random, and $0.0490$ for Interp-1BFGS. The selected MCTS-max and SSR-MCTS strategies are much worse, at $0.307$ and $0.314$, even though their best visited leaves have median condition errors $0.0526$ and $0.0603$. Thus the present data do not support a standalone MCTS advantage; they instead indicate that restriction/continuation and continuous refinement are the useful components in this regime.

The selected tree strategy is also much worse than the best leaf encountered on a run-by-run basis. Table~\ref{tab:mcts-gap-ci} reports bootstrap intervals for the median selected-minus-oracle gap. Because subtracting the same variational reference from both quantities leaves the gap unchanged, these intervals are identical whether expressed relative to the physical ground state or to the common variational reference.

\begin{table}[!htbp]
\centering
\caption{Selected-strategy minus oracle-best-visited normalized error for standalone MCTS. Values are the median over 100 runs (25 runs on each of four families), with percentile bootstrap 95\% intervals from 10,000 resamples.}
\label{tab:mcts-gap-ci}
\begin{tabular}{llrr}
\toprule
Observation & Method & 10,000 FEs & 30,000 FEs \\
\midrule
Exact & MCTS-max & $0.277\ [0.252,0.297]$ & $0.259\ [0.242,0.282]$ \\
Exact & SSR-MCTS & $0.251\ [0.215,0.275]$ & $0.241\ [0.217,0.290]$ \\
Low noise & MCTS-max & $0.271\ [0.242,0.296]$ & $0.258\ [0.227,0.270]$ \\
Low noise & SSR-MCTS & $0.250\ [0.219,0.261]$ & $0.266\ [0.250,0.289]$ \\
High noise & MCTS-max & $0.261\ [0.244,0.290]$ & $0.281\ [0.267,0.291]$ \\
High noise & SSR-MCTS & $0.250\ [0.226,0.272]$ & $0.265\ [0.244,0.287]$ \\
\bottomrule
\end{tabular}
\end{table}

\begin{figure}[!htbp]
\centering
\includegraphics[width=0.82\textwidth]{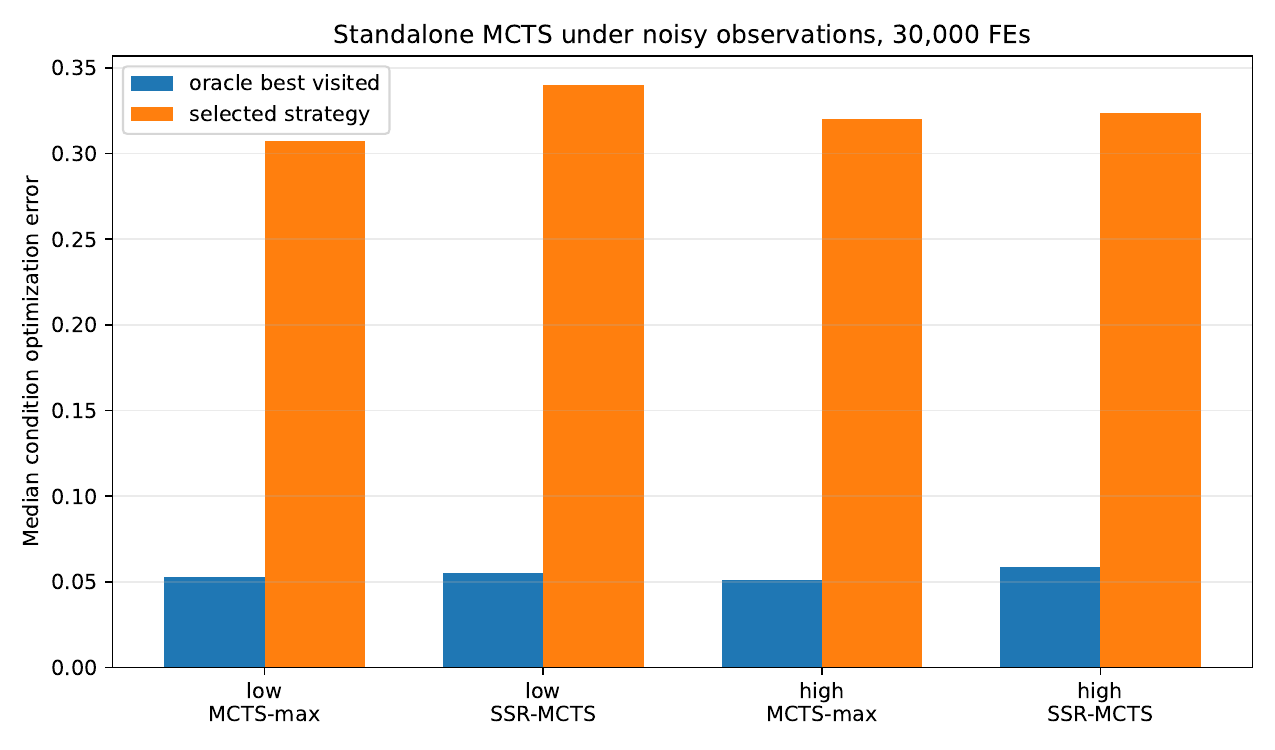}
\caption{Standalone MCTS under low and high additive observation noise at 30\,000 FEs, expressed as common-reference optimization error. Bars compare the best exact point visited with the exact quality of the strategy selected by the tree.}
\label{fig:mcts-noise-app}
\end{figure}

The tree diagnostics are treated only as supplementary mechanism descriptors. Prefix-based explained-variance measures can become large near the end of a short decision sequence because fine prefixes define small groups, and tree-distance correlations are not uniformly predictive across the tested families. The selected-versus-oracle outcome is therefore the primary MCTS diagnostic.

\section{Optimizer Hyperparameters and Reproducibility Details}
\label{app:hyperparameters}

All optimization routines operate over the hypercube $[-\pi, \pi]^D$, where $D = 2p = 6$ for the primary depth $p = 3$. Deterministic statevector evaluations and additive Gaussian observation-noise realizations are managed by a centralized objective evaluation wrapper with deterministic pseudo-random seeds. For population-based metaheuristics, a uniform population cap of $NP \le 150$ is enforced. At $D = 6$, this cap is inactive for all configured algorithms, ensuring that each baseline preserves its native parameter scaling without exhausting the function-evaluation (FE) budget on initialization.

The detailed hyperparameters, population-sizing rules, stopping criteria, and source implementations for all ten classical optimizers are summarized in Table~\ref{tab:hyperparameters}.

The settings were chosen according to three rules: library-native defaults were retained when they defined a complete baseline; literature/reference values were used for established algorithms such as SPSA and CEC-derived DE variants; and only budget-compatibility changes (population caps, restart wrappers, or the streamlined jSO implementation) were introduced uniformly across models. No values were tuned separately on Max-Cut, EA, diluted three-spin, SK, or either noise level. This avoids test-instance tuning leakage, but also means the study compares fixed implementations rather than fully tuned optimizer families.

The population sizes in Table~\ref{tab:hyperparameters} should therefore be read as part of the optimizer definitions rather than as a controlled nuisance parameter. In particular, L-SRTDE begins at $20D=120$ individuals in the primary $D=6$ study and then reduces the population toward four individuals, while iL-SHADE starts at 24. We impose the same FE ceiling rather than equalizing generations, so the broader L-SRTDE population simultaneously provides more contemporaneous samples and fewer generations for the same FE allowance. This distinction becomes especially relevant when interpreting noisy-objective robustness.

\begin{table}[!htbp]
\centering
\small
\caption{Hyperparameter settings, population configurations, and implementation details for the benchmarked optimizers ($D=6$, budgets $B \in \{10\,000, 30\,000\}$ FEs).}
\label{tab:hyperparameters}
\resizebox{\textwidth}{!}{%
\begin{tabular}{lp{4.2cm}p{4.8cm}p{3.2cm}}
\hline
\textbf{Optimizer} & \textbf{Population / Initialization} & \textbf{Key Hyperparameters} & \textbf{Termination / Source} \\
\hline
BFGS & Single random point $x_0 \sim \mathcal{U}(-\pi, \pi)^D$ & Numerical gradient step $h = 2\times 10^{-5}$ & $g_{\text{tol}} = 10^{-8}$, $\text{maxiter}=1800$; \texttt{scipy.optimize} \\
MS-BFGS & Repeated $x_0 \sim \mathcal{U}(-\pi, \pi)^D$; 1st start matched to BFGS & Independent BFGS restarts until FE budget is exhausted & Budget exhaustion ($B$); custom loop \\
SPSA & Single random point $x_0 \sim \mathcal{U}(-\pi, \pi)^D$ & $a=0.3$, $c=0.1$, $A=100.0$, $\alpha=0.602$, $\gamma=0.101$; 2-point Bernoulli perturbation & Budget exhaustion ($B$); custom implementation \\
CMA-ES & $NP = 4 + \lfloor 3\ln(D)\rfloor = 9$; initial $\sigma_0 = 1.0$ & Fully adaptive covariance matrix; diagonal bounds clipping & $\text{tol}_x=10^{-10}$, $\text{tol}_{\text{fun}}=10^{-10}$; \texttt{pycma} backend \\
MS-CMA-ES & $NP = 9$; 1st run matched to CMA-ES & Independent CMA-ES restarts launched after termination & Budget exhaustion ($B$); custom loop \\
iSOMA \cite{Diep2022} & $NP = \min(50, NP_{\text{cap}}) = 50$ individuals & $N_{\text{jump}}=10$, $\text{Step}=0.3$, $m=10$, $n=5$, $k=15$, $\text{PRT} = 0.1 + 0.9(FEs/\text{Max\_FEs})$ & Budget exhaustion ($B$); custom \texttt{iSOMA} package \\
SciPy-DE & $NP = \max(5, 2D) = 12$ individuals & Strategy: \texttt{currenttobest1bin}; $F \in (0.4, 0.9)$, $CR=0.8$, deferred updating & Polish disabled, $\text{tol}=0.0$; \texttt{scipy.optimize} \\
iL-SHADE & Initial $NP = \max(8, 4D) = 24$, linear reduction to $NP_{\text{min}}=4$ & External archive $A = 1.4 \cdot NP$; historical memory $H = 5$; dynamic $p_{\text{best}}$ & Budget exhaustion ($B$); \texttt{pyade.ilshade} \\
L-SRTDE & Initial $NP = 20D = 120$, linear reduction to $NP_{\text{min}}=4$ & Success-rate scaling: $\mu_F = 0.4 + 0.25\tanh(5\cdot SR)$, $\sigma_F=0.02$; memory $H=5$ for $Cr$ & Budget exhaustion ($B$); native CEC-2024 port \\
jSO-lite & Initial $NP = \max(30, \lfloor 25\sqrt{D}\log_{10}D\rfloor) = 47$, linear reduction to $NP_{\text{min}}=4$ & Dynamic $p_{\text{best}} \in [0.25, 0.125]$; memory $H=5$ for $F$ and $CR$ ($M_F=0.5, M_{CR}=0.8$) & Budget exhaustion ($B$); adapted CEC-2017 variant \\
\hline
\end{tabular}}
\end{table}

Standard jSO, introduced as the winner of the CEC 2017 benchmark competition, incorporates several specialized mechanisms designed for extended evaluation horizons (typically $10^4 \cdot D$ FEs). In this study, we deploy \textbf{jSO-lite}, a tailored variant configured for the tight evaluation budgets ($10\,000$ and $30\,000$ FEs) inherent to quantum circuit optimization:
\begin{itemize}
    \item \textbf{Mutation Strategy:} In place of the original weighted mutation operator (\textit{current-to-$p\text{best-}w/1$}) that uses individual-dependent scaling factor modifications ($F_w$), jSO-lite applies standard \textit{current-to-$p\text{best}/1$} mutation driven directly by sampled $F$ values.
    \item \textbf{Abolition of Stage-Dependent Memory Constraints:} Canonical jSO enforces rigid generation-dependent bounds on parameter adaptation (e.g., locking $F \le 0.7$ and $CR \in [0.5, 0.7]$ during the first third of the evaluations, and hardcoding the terminal memory slot to $M_F[H-1]=0.9, M_{CR}[H-1]=0.9$). jSO-lite discards these early-phase throttling rules, allowing memory cells to immediately adapt to the local landscape geometry via the weighted Lehmer mean.
    \item \textbf{Initial Population Sizing:} Standard jSO scales initial population size via $NP_{\text{init}} = \lfloor 25\sqrt{D}\log(D)\rfloor$, which yields excessively large initial populations in higher dimensions. jSO-lite uses $NP_{\rm init}=\max(30, \lfloor 25\sqrt{D}\log_{10}(D)\rfloor)$; at $D=6$, $25\sqrt{6}\log_{10}6\simeq47.65$, so the implemented integer population is $47$, not $30$, avoiding the expenditure of a substantial portion of the FE budget on non-adaptive uniform initialization.
    \item \textbf{Boundary Handling:} Boundary violations are addressed via coordinate-wise clamping onto $[-\pi, \pi]^D$ rather than periodic modulo arithmetic or random repositioning, ensuring consistent exploitation near the coordinate limits.
\end{itemize}
These modifications define the streamlined implementation used in the primary benchmark. We additionally ran the supplied paper-faithful jSO implementation on the same four selected representatives with the same run seeds, budgets, and observation-noise definitions. That control uses the original weighted mutation and stage constraints and an initial population of approximately 110 individuals at $D=6$. Table~\ref{tab:jso-canonical-ablation} shows that the modification has its clearest effect at the short exact budget: jSO-lite has an equal-family median oracle error $0.0068$ versus $0.0277$ for canonical jSO at 10\,000 FEs, and three of four family-wise paired Wilcoxon comparisons remain significant after Holm correction. At 30\,000 exact FEs both variants reach essentially the same numerical reference. Under low and high noise, no family-wise paired comparison survives Holm correction. We therefore retain jSO-lite as a budget-adapted variant and do not infer general superiority over canonical jSO.

\begin{table}[!htbp]
\centering
\small
\caption{Paper-faithful canonical jSO versus jSO-lite on the same four selected representatives. Values are equal-family medians of condition medians. Exact rows use oracle error (which is also the returned-quality comparison in the absence of observation noise); noisy rows additionally report selected error. Negative differences favor canonical jSO.}
\label{tab:jso-canonical-ablation}
\begin{tabular}{llrrrr}
\toprule
Noise & Budget & Canonical oracle & jSO-lite oracle & $\Delta_{\rm oracle}$ & $\Delta_{\rm selected}$\\
\midrule
high & 10,000 & 0.0516 & 0.0477 & +0.0026 & +0.0037 \\
high & 30,000 & 0.0301 & 0.0348 & -0.0034 & +0.0013 \\
low & 10,000 & 0.0336 & 0.0218 & +0.0072 & +0.0059 \\
low & 30,000 & 0.0055 & 0.0059 & +0.0021 & +0.0002 \\
none & 10,000 & 0.0277 & 0.0068 & +0.0166 & +0.0166 \\
none & 30,000 & 0.0000 & 0.0000 & -0.0000 & -0.0000 \\
\bottomrule
\end{tabular}
\end{table}

The realized FE counts are reported in Table~\ref{tab:fe-usage-audit}. Population and multistart methods use the full ceiling apart from SciPy-DE's generation-granularity shortfall at 10\,000 FEs, whereas single-start BFGS and CMA-ES frequently satisfy native termination criteria much earlier. These counts are part of the interpretation of the restart controls rather than an unreported resource difference.

\begin{table}[!htbp]
\centering
\scriptsize
\caption{Actual FE usage in the primary benchmark. Each cell is median [min,max] across the four selected families and 25 runs per family. Budgets are ceilings rather than forced evaluation counts; multistart and population methods generally exhaust them, whereas single-start BFGS and CMA-ES may terminate earlier.}
\label{tab:fe-usage-audit}
\resizebox{\textwidth}{!}{%
\begin{tabular}{lrrrrrr}
\toprule
& \multicolumn{2}{c}{Exact} & \multicolumn{2}{c}{Low noise} & \multicolumn{2}{c}{High noise}\\
\cmidrule(lr){2-3}\cmidrule(lr){4-5}\cmidrule(lr){6-7}
Optimizer & 10k & 30k & 10k & 30k & 10k & 30k\\
\midrule
BFGS & 556 [189,1,411] & 520 [182,1,098] & 180 [117,347] & 173 [117,314] & 180 [117,347] & 173 [117,314] \\
MS-BFGS & 10,000 & 30,000 & 10,000 & 30,000 & 10,000 & 30,000 \\
CMA-ES & 2,022 [1,378,5,185] & 2,040 [1,180,4,573] & 5,221 [4,546,9,541] & 5,311 [4,546,13,006] & 5,176 [4,546,9,991] & 5,244 [4,546,9,676] \\
MS-CMA-ES & 10,000 & 30,000 & 10,000 & 30,000 & 10,000 & 30,000 \\
SPSA & 10,000 & 30,000 & 10,000 & 30,000 & 10,000 & 30,000 \\
iSOMA & 10,000 & 30,000 & 10,000 & 30,000 & 10,000 & 30,000 \\
SciPy-DE & 9,996 & 30,000 & 9,996 & 30,000 & 9,996 & 30,000 \\
iL-SHADE & 10,000 & 30,000 & 10,000 & 30,000 & 10,000 & 30,000 \\
L-SRTDE & 10,000 & 30,000 & 10,000 & 30,000 & 10,000 & 30,000 \\
jSO-lite & 10,000 & 30,000 & 10,000 & 30,000 & 10,000 & 30,000 \\
\bottomrule
\end{tabular}}
\end{table}

A natural robustness extension is nested automated hyperparameter tuning. For example, Optuna-style Bayesian optimization could tune each optimizer on separate training instances/seeds, after which the chosen configuration would be frozen and evaluated on the held-out benchmark. The tuning evaluations must be accounted separately from the final test FE budgets; otherwise tuned methods would receive additional information not available to fixed baselines.

\clearpage

\scriptsize
\bibliographystyle{elsarticle-num}

\nocite{FarhiSK2022,Wang2021,Hansen2009Uncertainty,Shor1995,Novak2026LandscapeData}
\bibliography{references}

@article{Agirre2025MCTS,
  author  = {Agirre, A. and others},
  title   = {A Monte Carlo Tree Search approach to {QAOA}: finding a needle in the haystack},
  journal = {New Journal of Physics},
  volume  = {27},
  pages   = {043014},
  year    = {2025},
  doi     = {10.1088/1367-2630/adc765}
}

@misc{Farhi2014,
  author       = {Farhi, Edward and Goldstone, Jeffrey and Gutmann, Sam},
  title        = {A Quantum Approximate Optimization Algorithm},
  year         = {2014},
  eprint       = {1411.4028},
  archivePrefix= {arXiv}
}

@article{Zhou2020,
  author  = {Zhou, Leo and Wang, Sheng-Tao and Choi, Soonwon and Pichler, Hannes and Lukin, Mikhail D.},
  title   = {Quantum Approximate Optimization Algorithm: Performance, Mechanism, and Implementation on Near-Term Devices},
  journal = {Physical Review X},
  volume  = {10},
  pages   = {021067},
  year    = {2020},
  doi     = {10.1103/PhysRevX.10.021067}
}

@article{Willsch2020,
  author  = {Willsch, Madita and Willsch, Dennis and Jin, Fengping and De Raedt, Hans and Michielsen, Kristel},
  title   = {Benchmarking the quantum approximate optimization algorithm},
  journal = {Quantum Information Processing},
  volume  = {19},
  pages   = {197},
  year    = {2020},
  doi     = {10.1007/s11128-020-02692-8}
}

@article{SackSerbyn2021,
  author  = {Sack, Stefan H. and Serbyn, Maksym},
  title   = {Quantum annealing initialization of the quantum approximate optimization algorithm},
  journal = {Quantum},
  volume  = {5},
  pages   = {491},
  year    = {2021},
  doi     = {10.22331/q-2021-07-01-491}
}

@article{BoyWales2024,
  author  = {Boy, Charlie and Wales, David J.},
  title   = {Energy landscapes for the quantum approximate optimization algorithm},
  journal = {Physical Review A},
  volume  = {109},
  pages   = {062602},
  year    = {2024},
  doi     = {10.1103/PhysRevA.109.062602}
}

@inproceedings{Lavrijsen2020,
  author    = {Lavrijsen, Wim and Tudor, Ana and M{\"u}ller, Juliane and Iancu, Costin and de Jong, Wibe},
  title     = {Classical Optimizers for Noisy Intermediate-Scale Quantum Devices},
  booktitle = {2020 IEEE International Conference on Quantum Computing and Engineering (QCE)},
  pages     = {267--277},
  year      = {2020},
  doi       = {10.1109/QCE49297.2020.00041}
}

@article{FernandezPendas2022,
  author  = {Fern{\'a}ndez-Pend{\'a}s, Mario and Combarro, El{\'i}as F. and Vallecorsa, Sofia and Ranilla, Jos{\'e} and R{\'u}a, Ignacio F.},
  title   = {A study of the performance of classical minimizers in the Quantum Approximate Optimization Algorithm},
  journal = {Journal of Computational and Applied Mathematics},
  volume  = {404},
  pages   = {113388},
  year    = {2022},
  doi     = {10.1016/j.cam.2021.113388}
}

@article{PellowJarman2024,
  author  = {Pellow-Jarman, Alexander and McFarthing, Samuel and Sinayskiy, Ilya and Park, Daniel K. and Pillay, Anban and Petruccione, Francesco},
  title   = {The effect of classical optimizers and Ansatz depth on {QAOA} performance in noisy devices},
  journal = {Scientific Reports},
  volume  = {14},
  pages   = {16011},
  year    = {2024},
  doi     = {10.1038/s41598-024-66625-6}
}

@article{Wang2021,
  author  = {Wang, Samson and Fontana, Enrico and Cerezo, M. and Sharma, Kunal and Sone, Akira and Cincio, Lukasz and Coles, Patrick J.},
  title   = {Noise-induced barren plateaus in variational quantum algorithms},
  journal = {Nature Communications},
  volume  = {12},
  pages   = {6961},
  year    = {2021},
  doi     = {10.1038/s41467-021-27045-6}
}

@article{FarhiSK2022,
  author  = {Farhi, Edward and Goldstone, Jeffrey and Gutmann, Sam and Zhou, Leo},
  title   = {The Quantum Approximate Optimization Algorithm and the Sherrington--Kirkpatrick Model at Infinite Size},
  journal = {Quantum},
  volume  = {6},
  pages   = {759},
  year    = {2022},
  doi     = {10.22331/q-2022-07-07-759}
}

@article{Spall1992,
  author  = {Spall, James C.},
  title   = {Multivariate stochastic approximation using a simultaneous perturbation gradient approximation},
  journal = {IEEE Transactions on Automatic Control},
  volume  = {37},
  pages   = {332--341},
  year    = {1992},
  doi     = {10.1109/9.119632}
}

@incollection{Hansen2006,
  author    = {Hansen, Nikolaus},
  title     = {The {CMA} Evolution Strategy: A Comparing Review},
  booktitle = {Towards a New Evolutionary Computation},
  series    = {Studies in Fuzziness and Soft Computing},
  volume    = {192},
  pages     = {75--102},
  publisher = {Springer},
  year      = {2006},
  doi       = {10.1007/3-540-32494-1_4}
}

@article{StornPrice1997,
  author  = {Storn, Rainer and Price, Kenneth},
  title   = {Differential Evolution--A Simple and Efficient Heuristic for Global Optimization over Continuous Spaces},
  journal = {Journal of Global Optimization},
  volume  = {11},
  pages   = {341--359},
  year    = {1997},
  doi     = {10.1023/A:1008202821328}
}

@inproceedings{Brest2016,
  author    = {Brest, Janez and Mau{\v c}ec, Mirjam Sepesy and Bo{\v s}kovi{\'c}, Borko},
  title     = {{iL-SHADE}: Improved {L-SHADE} algorithm for single objective real-parameter optimization},
  booktitle = {2016 IEEE Congress on Evolutionary Computation (CEC)},
  pages     = {1188--1195},
  year      = {2016},
  doi       = {10.1109/CEC.2016.7743922}
}

@inproceedings{Brest2017,
  author    = {Brest, Janez and Mau{\v c}ec, Mirjam Sepesy and Bo{\v s}kovi{\'c}, Borko},
  title     = {Single objective real-parameter optimization: Algorithm {jSO}},
  booktitle = {2017 IEEE Congress on Evolutionary Computation (CEC)},
  pages     = {1311--1318},
  year      = {2017},
  doi       = {10.1109/CEC.2017.7969456}
}

@inproceedings{Stanovov2024,
  author    = {Stanovov, Vladimir and Semenkin, Eugene},
  title     = {Success Rate-based Adaptive Differential Evolution {L-SRTDE} for {CEC} 2024 Competition},
  booktitle = {2024 IEEE Congress on Evolutionary Computation (CEC)},
  pages     = {1--8},
  year      = {2024},
  doi       = {10.1109/CEC60901.2024.10611907}
}

@article{BonetMonroig2023,
  author  = {Bonet-Monroig, Xavier and Wang, Hao and Vermetten, Diederick and Senjean, Bruno and Moussa, Charles and B{\"a}ck, Thomas and Dunjko, Vedran and O'Brien, Thomas E.},
  title   = {Performance comparison of optimization methods on variational quantum algorithms},
  journal = {Physical Review A},
  volume  = {107},
  pages   = {032407},
  year    = {2023},
  doi     = {10.1103/PhysRevA.107.032407}
}

@article{Novak2026CEC,
  author  = {Nov{\'a}k, Vojt{\v e}ch and Bezd{\v e}k, Tom{\'a}{\v s} and Zelinka, Ivan and Das, Swagatam and Beseda, Martin},
  title   = {A longitudinal analysis of the {CEC} single-objective competitions (2010--2024) and implications for variational quantum optimization},
  journal = {Swarm and Evolutionary Computation},
  volume  = {107},
  pages   = {102469},
  year    = {2026},
  doi     = {10.1016/j.swevo.2026.102469}
}

@article{Piotrowski2026,
  author  = {Piotrowski, Adam P. and Piotrowska, Anna E. and Napiorkowski, Jaroslaw J.},
  title   = {Experimental survey of {L-SHADE} and {SHADE}-based adaptive differential evolution algorithms},
  journal = {Swarm and Evolutionary Computation},
  volume  = {101},
  pages   = {102286},
  year    = {2026},
  doi     = {10.1016/j.swevo.2026.102286}
}

@article{Diep2022,
  author  = {Diep, Q. B. and Truong, T. C. and Das, Swagatam and Zelinka, Ivan},
  title   = {Self-Organizing Migrating Algorithm with narrowing search space strategy for robot path planning},
  journal = {Applied Soft Computing},
  volume  = {116},
  pages   = {108270},
  year    = {2022},
  doi     = {10.1016/j.asoc.2021.108270}
}

@article{Rakshit2017,
  author  = {Rakshit, Pratyusha and Konar, Amit and Das, Swagatam},
  title   = {Noisy evolutionary optimization algorithms--A comprehensive survey},
  journal = {Swarm and Evolutionary Computation},
  volume  = {33},
  pages   = {18--45},
  year    = {2017},
  doi     = {10.1016/j.swevo.2016.09.002}
}

@article{Hansen2009Uncertainty,
  author  = {Hansen, Nikolaus and Niederberger, A. S. P. and Guzzella, Lino and Koumoutsakos, Petros},
  title   = {A method for handling uncertainty in evolutionary optimization with an application to feedback control of combustion},
  journal = {IEEE Transactions on Evolutionary Computation},
  volume  = {13},
  pages   = {180--197},
  year    = {2009},
  doi     = {10.1109/TEVC.2008.924423}
}

@article{ArnoldBeyer2002,
  author  = {Arnold, Dirk V. and Beyer, Hans-Georg},
  title   = {Performance analysis of evolution strategies with multi-recombination in high-dimensional {$\mathbb{R}^N$} search spaces disturbed by noise},
  journal = {Theoretical Computer Science},
  volume  = {289},
  pages   = {629--647},
  year    = {2002},
  doi     = {10.1016/S0304-3975(01)00384-X}
}

@article{ArnoldBeyer2003,
  author  = {Arnold, Dirk V. and Beyer, Hans-Georg},
  title   = {On the benefits of populations for noisy optimization},
  journal = {Evolutionary Computation},
  volume  = {11},
  pages   = {111--127},
  year    = {2003},
  doi     = {10.1162/106365603766646799}
}

@article{NishidaAkimoto2017,
  author  = {Nishida, K. and Akimoto, Y.},
  title   = {Proposal and evaluation of the population size adaptation for the covariance matrix adaptation evolution strategy},
  journal = {Transactions of the Japanese Society for Evolutionary Computation},
  volume  = {8},
  pages   = {61--74},
  year    = {2017},
  doi     = {10.11394/tjpnsec.8.61}
}

@article{Preskill2018,
  author  = {Preskill, John},
  title   = {Quantum Computing in the {NISQ} era and beyond},
  journal = {Quantum},
  volume  = {2},
  pages   = {79},
  year    = {2018},
  doi     = {10.22331/q-2018-08-06-79}
}

@article{Kandala2017,
  author  = {Kandala, Abhinav and Mezzacapo, Antonio and Temme, Kristan and Takita, Maika and Brink, Markus and Chow, Jerry M. and Gambetta, Jay M.},
  title   = {Hardware-efficient variational quantum eigensolver for small molecules and quantum magnets},
  journal = {Nature},
  volume  = {549},
  pages   = {242--246},
  year    = {2017},
  doi     = {10.1038/nature23879}
}

@article{Shor1995,
  author  = {Shor, Peter W.},
  title   = {Scheme for reducing decoherence in quantum computer memory},
  journal = {Physical Review A},
  volume  = {52},
  pages   = {R2493--R2496},
  year    = {1995},
  doi     = {10.1103/PhysRevA.52.R2493}
}

@article{Wang2021NoiseBP,
  author  = {Wang, Samson and Fontana, Enrico and Cerezo, M. and Sharma, Kunal and Sone, Akira and Cincio, Lukasz and Coles, Patrick J.},
  title   = {Noise-induced barren plateaus in variational quantum algorithms},
  journal = {Nature Communications},
  volume  = {12},
  pages   = {6961},
  year    = {2021},
  doi     = {10.1038/s41467-021-27045-6}
}

@article{Cerezo2021Review,
  author  = {Cerezo, M. and Arrasmith, Andrew and Babbush, Ryan and Benjamin, Simon C. and Endo, Suguru and Fujii, Keisuke and McClean, Jarrod R. and Mitarai, Kosuke and Yuan, Xiao and Cincio, Lukasz and Coles, Patrick J.},
  title   = {Variational quantum algorithms},
  journal = {Nature Reviews Physics},
  volume  = {3},
  pages   = {625--644},
  year    = {2021},
  doi     = {10.1038/s42254-021-00348-9}
}

@article{Tilly2022,
  author  = {Tilly, Jules and Chen, Hongxiang and Cao, Shuxiang and Picozzi, Dario and Setia, Kanav and Li, Ying and Grant, Edward and Wossnig, Leonard and Rungger, Ivan and Booth, George H. and Tennyson, Jonathan},
  title   = {The Variational Quantum Eigensolver: A review of methods and best practices},
  journal = {Physics Reports},
  volume  = {986},
  pages   = {1--128},
  year    = {2022},
  doi     = {10.1016/j.physrep.2022.08.003}
}

@article{AnschuetzKiani2022,
  author  = {Anschuetz, Eric R. and Kiani, Bobak T.},
  title   = {Quantum variational algorithms are swamped with traps},
  journal = {Nature Communications},
  volume  = {13},
  pages   = {7760},
  year    = {2022},
  doi     = {10.1038/s41467-022-35364-5}
}

@article{Nemkov2025,
  author  = {Nemkov, Nikita A. and Kiktenko, Evgeniy O. and Fedorov, Alexey K.},
  title   = {Barren plateaus swamped with traps},
  journal = {Physical Review A},
  volume  = {111},
  pages   = {012441},
  year    = {2025},
  doi     = {10.1103/PhysRevA.111.012441}
}

@article{Mele2022,
  author  = {Mele, Antonio Anna and Mbeng, Glen Bigan and Santoro, Giuseppe E. and Collura, Mario and Torta, Paola},
  title   = {Avoiding barren plateaus via transferability of smooth solutions in a Hamiltonian variational ansatz},
  journal = {Physical Review A},
  volume  = {106},
  pages   = {L060401},
  year    = {2022},
  doi     = {10.1103/PhysRevA.106.L060401}
}

@article{Failde2023,
  author  = {Fa{\'i}lde, D. and Viqueira, J. D. and Mussa Juane, M. and G{\'o}mez, A.},
  title   = {Using Differential Evolution to avoid local minima in Variational Quantum Algorithms},
  journal = {Scientific Reports},
  volume  = {13},
  pages   = {16230},
  year    = {2023},
  doi     = {10.1038/s41598-023-43404-3}
}

@article{Scriva2024,
  author  = {Scriva, G. and Astrakhantsev, N. and Pilati, S. and Mazzola, G.},
  title   = {Challenges of variational quantum optimization with measurement shot noise},
  journal = {Physical Review A},
  volume  = {109},
  pages   = {032408},
  year    = {2024},
  doi     = {10.1103/PhysRevA.109.032408}
}

@article{Jones2025,
  author  = {Jones, B. D. M. and Mineh, L. and Montanaro, A.},
  title   = {Benchmarking a wide range of optimisers for solving the Fermi--Hubbard model using the variational quantum eigensolver},
  journal = {Quantum Science and Technology},
  volume  = {10},
  pages   = {045032},
  year    = {2025},
  doi     = {10.1088/2058-9565/adfe15}
}

@article{Cheng2024DARBO,
  author  = {Cheng, L. and Chen, Y.-Q. and Zhang, S.-X. and Zhang, S.},
  title   = {Quantum approximate optimization via learning-based adaptive optimization},
  journal = {Communications Physics},
  volume  = {7},
  pages   = {83},
  year    = {2024},
  doi     = {10.1038/s42005-024-01577-x}
}

@article{Lyngfelt2025Transfer,
  author  = {Lyngfelt, I. and Garc{\'i}a-{\'A}lvarez, L.},
  title   = {Symmetry-informed transferability of optimal parameters in the quantum approximate optimization algorithm},
  journal = {Physical Review A},
  volume  = {111},
  pages   = {022418},
  year    = {2025},
  doi     = {10.1103/PhysRevA.111.022418}
}

@inproceedings{TanabeFukunaga2014,
  author    = {Tanabe, Ryoji and Fukunaga, Alex S.},
  title     = {Improving the Search Performance of {SHADE} Using Linear Population Size Reduction},
  booktitle = {2014 IEEE Congress on Evolutionary Computation (CEC)},
  pages     = {1658--1665},
  year      = {2014},
  doi       = {10.1109/CEC.2014.6900380}
}

@article{Piotrowski2023,
  author  = {Piotrowski, Adam P. and Napiorkowski, Jaroslaw J. and Piotrowska, Anna E.},
  title   = {Choice of benchmark optimization problems does matter},
  journal = {Swarm and Evolutionary Computation},
  volume  = {83},
  pages   = {101378},
  year    = {2023},
  doi     = {10.1016/j.swevo.2023.101378}
}

@article{LaTorre2021,
  author  = {LaTorre, Antonio and Molina, Daniel and Osaba, Eneko and Poyatos, Jon and Del Ser, Javier and Herrera, Francisco},
  title   = {A prescription of methodological guidelines for comparing bio-inspired optimization algorithms},
  journal = {Swarm and Evolutionary Computation},
  volume  = {67},
  pages   = {100973},
  year    = {2021},
  doi     = {10.1016/j.swevo.2021.100973}
}

@article{Novak2025NoisyLandscapes,
  author  = {Nov{\'a}k, Vojt{\v e}ch and Zelinka, Ivan and Sn{\'a}{\v s}el, V{\'a}clav},
  title   = {Optimization strategies for variational quantum algorithms in noisy landscapes},
  journal = {Evolutionary Intelligence},
  volume  = {19},
  pages   = {142},
  year    = {2026},
  doi     = {10.1007/s12065-026-01248-6}
}

@misc{Illesova2025VHA,
  author        = {Ill{\'e}sov{\'a}, Silvie and Nov{\'a}k, Vojt{\v e}ch and Bezd{\v e}k, Tom{\'a}{\v s} and Possel, C. and Beseda, Martin},
  title         = {Numerical Optimization Strategies for the Variational Hamiltonian Ansatz in Noisy Quantum Environments},
  year          = {2025},
  eprint        = {2505.22398},
  archivePrefix = {arXiv},
  primaryClass  = {quant-ph}
}

@misc{Illesova2025Statistical,
  author        = {Ill{\'e}sov{\'a}, Silvie and Bezd{\v e}k, Tom{\'a}{\v s} and Nov{\'a}k, Vojt{\v e}ch and Senjean, Bruno and Beseda, Martin},
  title         = {Statistical Benchmarking of Optimization Methods for Variational Quantum Eigensolver under Quantum Noise},
  year          = {2025},
  eprint        = {2510.08727},
  archivePrefix = {arXiv},
  primaryClass  = {quant-ph}
}

@misc{Bezdek2025ClassicalOptimization,
  author        = {Bezd{\v e}k, Tom{\'a}{\v s} and Yuan, H. and Nov{\'a}k, Vojt{\v e}ch and Ill{\'e}sov{\'a}, Silvie and Beseda, Martin},
  title         = {Classical Optimization Strategies for Variational Quantum Algorithms: A Systematic Study of Noise Effects and Parameter Efficiency},
  year          = {2025},
  eprint        = {2511.09314},
  archivePrefix = {arXiv},
  primaryClass  = {quant-ph}
}

@misc{Novak2025Reliable,
  author        = {Nov{\'a}k, Vojt{\v e}ch and Ill{\'e}sov{\'a}, Silvie and Bezd{\v e}k, Tom{\'a}{\v s} and Zelinka, Ivan and Beseda, Martin},
  title         = {Reliable Optimization Under Noise in Quantum Variational Algorithms},
  year          = {2025},
  eprint        = {2511.08289},
  archivePrefix = {arXiv},
  primaryClass  = {quant-ph}
}

@inproceedings{illesova2025qmetric,
  author    = {Ill{\'e}sov{\'a}, Silvie and Rybotycki, Tomasz and Beseda, Martin},
  title     = {{QMetric}: Benchmarking Quantum Neural Networks Across Circuits, Features, and Training Dimensions},
  booktitle = {Proceedings of QualITA 2025: The Fourth Conference on System and Service Quality},
  series    = {CEUR Workshop Proceedings},
  volume    = {4080},
  year      = {2025}
}

@article{beseda2024state,
  author  = {Beseda, Martin and Ill{\'e}sov{\'a}, Silvie and Yalouz, Saad and Senjean, Bruno},
  title   = {State-Averaged Orbital-Optimized {VQE}: A quantum algorithm for the democratic description of ground and excited electronic states},
  journal = {Journal of Open Source Software},
  volume  = {9},
  pages   = {6036},
  year    = {2024},
  doi     = {10.21105/joss.06036}
}

@article{illesova2025transformation,
  author  = {Ill{\'e}sov{\'a}, Silvie and Beseda, Martin and Yalouz, Saad and Lasorne, Benjamin and Senjean, Bruno},
  title   = {Transformation-Free Generation of a Quasi-Diabatic Representation from the State-Average Orbital-Optimized Variational Quantum Eigensolver},
  journal = {Journal of Chemical Theory and Computation},
  volume  = {21},
  pages   = {5457--5480},
  year    = {2025},
  doi     = {10.1021/acs.jctc.5c00327}
}

@misc{Novak2026Landscape,
  author        = {Nov{\'a}k, Vojt{\v e}ch and Zelinka, Ivan and Das, Swagatam and Beseda, Martin},
  title         = {Optimization Landscape Geometry in {VQE} for Frustrated Quantum Spin Models},
  year          = {2026},
  eprint        = {2609.00235},
  archivePrefix = {arXiv},
  primaryClass  = {quant-ph},
  doi           = {10.48550/arXiv.2609.00235}
}

@misc{Novak2026LandscapeData,
  author    = {Nov{\'a}k, Vojt{\v e}ch and Zelinka, Ivan and Das, Swagatam and Beseda, Martin},
  title     = {Optimization Landscape Geometry in {VQE} for Frustrated Quantum Spin Models},
  note      = {Dataset, Zenodo},
  year      = {2026},
  doi       = {10.5281/zenodo.22210896}
}

@misc{Novak2026GlobalSearch,
  author        = {Nov{\'a}k, Vojt{\v e}ch and Zelinka, Ivan},
  title         = {When is global evolutionary search useful for variational quantum algorithms? A landscape-first study},
  year          = {2026},
  eprint        = {2609.14594},
  archivePrefix = {arXiv},
  primaryClass  = {quant-ph}
}

@misc{Novak2026LinearProposal,
  author        = {Nov{\'a}k, Vojt{\v e}ch and Zelinka, Ivan},
  title         = {Linear Proposal Operators and Stochastic Search Geometry in {SOMA} and Differential Evolution},
  year          = {2026},
  eprint        = {2607.29228},
  archivePrefix = {arXiv},
  primaryClass  = {cs.NE}
}

@article{EdwardsAnderson1975,
  author  = {Edwards, S. F. and Anderson, P. W.},
  title   = {Theory of spin glasses},
  journal = {Journal of Physics F: Metal Physics},
  volume  = {5},
  pages   = {965--974},
  year    = {1975},
  doi     = {10.1088/0305-4608/5/5/017}
}

@article{GrossMezard1984,
  author  = {Gross, D. J. and M{\'e}zard, M.},
  title   = {The simplest spin glass},
  journal = {Nuclear Physics B},
  volume  = {240},
  pages   = {431--452},
  year    = {1984},
  doi     = {10.1016/0550-3213(84)90237-2}
}

@article{FranzParisiTriton1999,
  author  = {Franz, S. and M{\'e}zard, M. and Parisi, G. and Peliti, L.},
  title   = {Measuring equilibrium properties in disordered systems: A review of the replica and cavity methods},
  journal = {Journal of Statistical Physics},
  volume  = {97},
  pages   = {459--488},
  year    = {1999},
  doi     = {10.1023/A:1004602906332}
}

@article{SherringtonKirkpatrick1975,
  author  = {Sherrington, D. and Kirkpatrick, S.},
  title   = {Solvable Model of a Spin-Glass},
  journal = {Physical Review Letters},
  volume  = {35},
  pages   = {1792--1796},
  year    = {1975},
  doi     = {10.1103/PhysRevLett.35.1792}
}

@article{Farhi2022SKInfinite,
  author  = {Farhi, Edward and Goldstone, Jeffrey and Gutmann, Sam and Zhou, Leo},
  title   = {The Quantum Approximate Optimization Algorithm and the Sherrington--Kirkpatrick Model at Infinite Size},
  journal = {Quantum},
  volume  = {6},
  pages   = {759},
  year    = {2022},
  doi     = {10.22331/q-2022-07-07-759}
}

@article{novak2026quantum,
  title     = {Quantum machine learning for predicting anastomotic leak: a clinical study},
  author    = {Nov{\'a}k, Vojt{\v e}ch and Zelinka, Ivan and P{\v r}ibylov{\'a}, Lenka and Mart{\'\i}nek, Lubom{\'\i}r and Ben{\v c}urik, Vladim{\'\i}r},
  journal   = {Sci. Rep.},
  publisher = {Springer Science and Business Media LLC},
  volume    = {16},
  number    = {1},
  month     = may,
  year      = {2026},
  copyright = {https://creativecommons.org/licenses/by/4.0},
  language  = {en}
}

@article{illesova2025classical,
  title   = {From Classical to Hybrid: A Practical Framework for Quantum-Enhanced Learning},
  author  = {Ill{\'e}sov{\'a}, Silvie and Bezd{\v{e}}k, Tom{\'a}{\v{s}} and Nov{\'a}k, Vojt{\v{e}}ch and Zelinka, Ivan and Cacciatore, Stefano and Beseda, Martin},
  journal = {arXiv preprint arXiv:2511.08205},
  year    = {2025}
}

@article{illesova2025complementarity,
  title     = {On the Complementarity of Classical Convolution and Quantum Neural Networks in Image Classification},
  author    = {Ill{\'e}sov{\'a}, Silvie and Obeng, Emmanuel and Bezd{\v{e}}k, Tom{\'a}{\v{s}} and Nov{\'a}k, Vojt{\v{e}}ch and Beseda, Martin},
  year      = {2025},
  publisher = {Preprints},
  journal = {Preprints}
}

@article{illesova2026importance,
  title     = {On the importance of fundamental properties in quantum-classical machine learning models},
  author    = {Ill{\'e}sov{\'a}, Silvie and Rybotycki, Tomasz and Gawron, Piotr and Beseda, Martin},
  journal   = {International Journal of Parallel, Emergent and Distributed Systems},
  pages     = {1--28},
  year      = {2026},
  publisher = {Taylor \& Francis}
}

\end{document}